\documentclass[aps,showkeys,superscriptaddress,preprint,tightenlines]{revtex4}
\usepackage[colorlinks,linkcolor=blue,anchorcolor=blue,citecolor=blue]{hyperref}
\usepackage{times}
\usepackage{amsmath}
\usepackage{graphicx}
\usepackage{epsfig}
\usepackage{dcolumn}
\usepackage{bm}
\usepackage{color}
\usepackage{longtable}
\usepackage{array}
\usepackage{multirow}
\usepackage{subfigure}

\definecolor{gongnvlan}{rgb}{0.15,0.39,0.46}

\newcommand{\EE}{e^+e^-}

\newcommand{\br}{\mathcal{B}}
\newcommand{\BR}{\mathcal{B}}

\newcommand{\mbc}{M_\mathrm{BC}}
\newcommand{\ebeam}{E_\mathrm{beam}}

\newcommand{\pvect}{\vec{p}_\mathrm{tag}}

\newcommand{\mmiss}{M_\mathrm{miss}}

\newcommand{\egeg} {{\it e.g.}}
\newcommand{\ieie} {{\it i.e.}}

\newcommand{\lambdacp}{\Lambda_{c}^{+}}
\newcommand{\lambdacm}{\bar{\Lambda}{}_{c}^{-}}

\begin{document}

\title{\bf \boldmath
Study of the Standard Model with weak decays of charmed hadrons at BESIII}
\author{Hai-Bo Li} \email{lihb@ihep.ac.cn}

\affiliation{
    Institute of High Energy Physics, Beijing 100049, People's Republic of China
}
\affiliation{
    University of Chinese Academy of Sciences,  Beijing 100049, People's Republic of China
}
\author{Xiao-Rui Lyu} \email{xiaorui@ucas.ac.cn}
\affiliation{
    University of Chinese Academy of Sciences,  Beijing 100049, People's Republic of China
}


\begin{abstract}
   
   A comprehensive review of weak decays of charmed hadrons ($D^{0/+}$, $D^+_s$ and $\Lambda^+_c$) based on analyses of the threshold data from $e^+e^-$ annihilation in the BESIII experiment is presented .    
   Current  experimental challenges and successes in understanding  decays of the charmed hadrons are discussed.   Precise calibrations of QCD and tests of the Standard Model are provided by measurements of purely leptonic and semi-leptonic decays of charmed hadrons, and lepton universality  is probed in purely leptonic decays  of charmed mesons to  three generations of leptons. 
   Quantum correlations in threshold data samples provide access to strong phases in the neutral $D$ meson decays and  probe  the decay dynamics of the charmed $\Lambda_c$ baryon.     
    Charm physics studies with near-threshold production of charmed particle pairs are unique to BESIII, and provide many important opportunities and challenges.
 
   \end{abstract}
 

\maketitle
\oddsidemargin  -0.2cm    
\evensidemargin -0.2cm

\section{Introduction}

The discovery of the $J/\psi$  in 1974 marked a new era in particle physics.  The arrival  of the first heavy quark  indicated  that the Standard Model (SM) provided a correct low-energy description of particle physics.  Four decades later, the charmed quark still plays unique roles in studies of strong and weak interactions~\cite{Cerri:2018ypt}.   Recent observation of $CP$ violation in charmed meson decays has attracted significant and renewed interest to the charm physics~\cite{Aaij:2019kcg,Saur:2020rgd}. 
It paves the road to precise tests of SM in interesting weak interaction transitions 
and maybe even to searches for new physics beyond the SM.

\begin{figure}[hbp]
\centering
\includegraphics[width=1.3in]{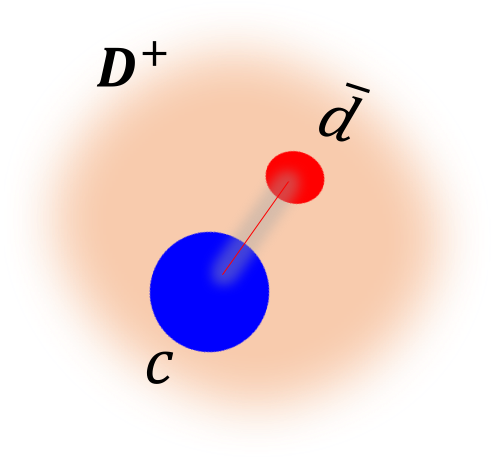}
\includegraphics[width=1.3in]{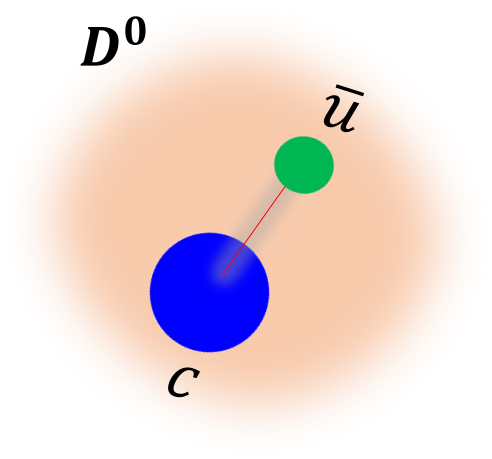}
\includegraphics[width=1.3in]{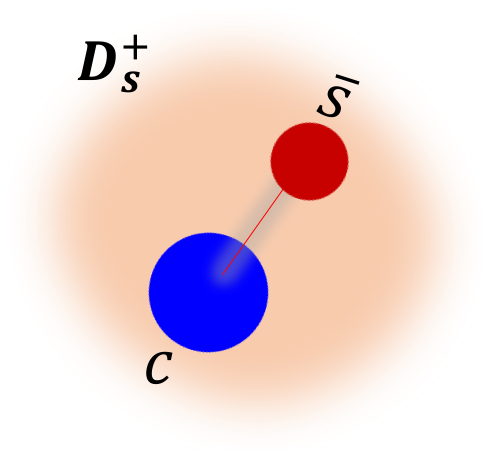}
\includegraphics[width=1.3in]{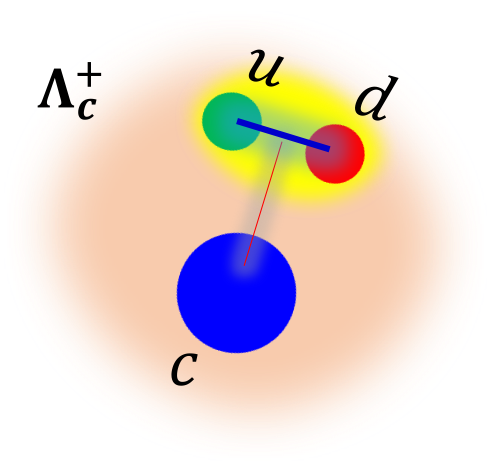}
\caption{\small Quark constituents for the ground-state charmed hadrons of $D^+$($c\bar{d}$), $D^0$($c\bar{u}$), $D_s^+$($c\bar{s}$) and  $\Lambda_c^+$($cud$). Taken from Ref.~\cite{physics:bes}.}
\label{fig:quark}
\end{figure}

A distinctive feature of all of the charmed hadrons is that their
masses place them at the edge of the region where non-perturbative hadronic physics is operative, forcing us to develop new
means to cope with such scales.  This point has been made in prescient reviews~\cite{Bianco:2003vb,Artuso:2008vf} that posed many of the questions that are still awaiting  answers.  
 While
this fact does not markedly affect the theoretical description of leptonic and semi-leptonic decays of charmed hadrons, it poses challenges to analyses of their hadronic transitions. We expect that detailed 
experimental studies  would provide some hints on the dynamics of charm hadronic decays, so that eventually those problems will 
be overcome.  
In this review we focus on the weak decays of ground-state charmed hadrons, \ieie, three charmed mesons $D^+$($c\bar{d}$), $D^0$($c\bar{u}$) and $D_s^+$($c\bar{s}$) as well as one charmed baryon $\Lambda_c^+$($cud$), with internal quark constituents as depicted in Fig.~\ref{fig:quark}, that can be extensively studied using data collected at the BESIII experiment. There are mainly  three classes of charmed hadron decays: purely leptonic, semi-leptonic,
and hadronic decays. Measurements of the charmed hadron decays can be used to calibrate lattice quantum chromodynamics (LQCD) calculations.  In addition,  BESIII data  provide stringent
constraints on the Cabibbo-Kobayashi-Maskawa (CKM)  six-quark flavour-mixing matrix~\cite{Kobayashi:1973fv} via: 1) precision measurements of the CKM matrix elements $|V_{cs}|$ and $|V_{cd}|$  that parametrize the strengths
of $c\to s$ and $c \to d$ weak transitions, respectively; 2) determinations of the strong-interaction phases in $D$-meson decays that are  essential inputs to  measurements of the $CP$-violating phase $\gamma$ of the CKM matrix element $V_{ub}$ in $B$-meson decays~\cite{Ceccucci:2020cim}.  

\section{Advantages near threshold production from $e^+e^-$ annihilation  } 

 Experiments at $e^+e^-$  machines operating at the $\psi(3770)$ and $\psi(4140)$ resonances and $\Lambda^+_c \bar{\Lambda}^-_c$ threshold, such as CLEO-c and BESIII~\cite{Ablikim:2019hff}, have several important advantages.
 First, the cross section for charm production is relatively high,  for example, $\sigma(e^+e^- \to D^0\bar{D}^0) = (3.615\pm 0.010 \pm 0.038)$ nb and $\sigma(e^+e^- \to D^+ D^-) = (2.830 \pm 0.011 \pm 0.026)$ nb at the $\psi(3770)$ peak~\cite{Ablikim:2018tay}. 
Second,  the $D\bar{D}$ and $\Lambda^+_c \bar{\Lambda}^-_c$ pairs are produced in the exclusive two-body channel with no additional particles.  
Thus, one can employ a double-tag technique pioneered by the Mark III experiment~\cite{Baltrusaitis:1985iw}, a full reconstruction of an anti-$D$ meson on one side of tagged events together with the known
momentum and energy of colliding beams provides a ``beam"  of $D$ particles of known four-momentum on the
other side. The tag yield, which provides the normalization for the branching fraction
measurement,  is extracted from the distribution of beam-constrained mass $\mbc  = \sqrt{\ebeam^2 - |\pvect |^2}$, where $\pvect$ is the three-momentum of the tag $\bar{D}$ candidate and $\ebeam$ is the beam energy , both evaluated in the $\EE$ center-of-mass system.  When a tagged $D^+$  decays
to a muon and a muonic neutrino, $\mu^+\nu_\mu$,  the mass of the (missing) nearly zero-mass neutrino
can be inferred from energy-momentum conservation. This tagging technique, which obviates the need for knowledge of the luminosity or the production cross section, is a powerful tool for charmed particle
decay measurements that is most accurately performed by the near-threshold experiments.  

 Furthermore,  the charmed hadron pairs at BESIII are produced via $e^+e^-$ annihilation through a virtual
photon (with spin, parity and $C$-parity of  $J^{PC} = 1^{--}$),   {\it e.g.,}  in the process $e^+e^- \to \gamma^*  \to \psi(3770) \to D^0\bar{D}^0$ ($e^+e^- \to \gamma^* \to \Lambda^+_c \bar{\Lambda}^-_c$). Hence the wave function of the produced charm hadron pairs is analogous to that of photons in an aligned, spin-1 state with odd charge parity $C=-1$,  and the $D^0\bar{D}^0$ ($\Lambda^+_c \bar{\Lambda}^-_c$) pair are in a quantum-entangled state.  This allows for unique probes of the structure of decay amplitudes and relative phases between $D^0$ and $\bar{D}^0$ decays, 
 as well as novel measurements of neutral $D$ mixing and $CP$ violation in $D^0$ and $\Lambda^+_c$ decays~\cite{Xing:1996pn, Kang:2010td, Kang:2009iy,Charles:2009ig}. 

\section{Precision Tests of the Standard Model}

In the SM,  quark-flavor mixing is characterized  by the unitary $3\times 3$ CKM matrix~\cite{Kobayashi:1973fv}: 
   \begin{equation}
V_{\rm CKM}
= \left(
\begin{array}{ccc}
 V_{ud} &  V_{us} &  V_{ub} \\
 V_{cd} &  V_{cs} &  V_{cb} \\
 V_{td} &  V_{ts} &  V_{tb}  \\
\end{array} \right ).
\label{eq:ckm-m}
\end{equation}
  The CKM matrix induces flavor-changing transitions within and among generations in the charged currents in tree-level $W^\pm$-exchange interactions. Experiments have revealed  a strong hierarchy among the CKM matrix elements: transitions within the same generation are described  by $V_{\rm CKM}$  elements of $\mathcal{O}(1)$,   whereas there is a suppression
of $\mathcal{O}(10^{-1})$ for transitions between the first and the second generations, $\mathcal{O}(10^{-2})$ between  the second and the third, and $\mathcal{O}(10^{-3})$ between
the first and the third.  Following the observation of this hierarchy,  Wolfenstein~\cite{Wolfenstein:1983yz} proposed an expansion of the 
CKM matrix in terms of four parameters (which was further modified by Buras~\cite{Buras:1994ec} ), $\lambda$, $A$, $\bar{\rho}$, and $\bar{\eta}$, under the following relations:
  \begin{equation}
\label{eq:lambda:A:rhoeta}
\lambda^2 =\frac{|V_{us}|^2}{|V_{ud}|^2+|V_{us}|^2}\,, \qquad
A^2\lambda^4=\frac{|V_{cb}|^2}{|V_{ud}|^2+|V_{us}|^2}\,, \qquad
\bar\rho+i\bar\eta=-\frac{V_{ud}V_{ub}^*}{V_{cd}V^*_{cb}},
\end{equation}
which are used to fully characterize the matrix.     
  Any deviation of $V_{\rm CKM}$ from unitarity would indicate new physics beyond the SM.  Therefore, improving our knowledge of the CKM matrix elements to test unitarity is one of the principal goals of flavor physics.  BESIII data  provide direct precise measurements of the CKM matrix elements $|V_{cs}|$ and $|V_{cd}|$ using the purely leptonic and semi-leptonic charmed-hadron decay rates  as discussed in detail below. 

\begin{figure}[hbp]
\centering
    \includegraphics[width=6.2in]{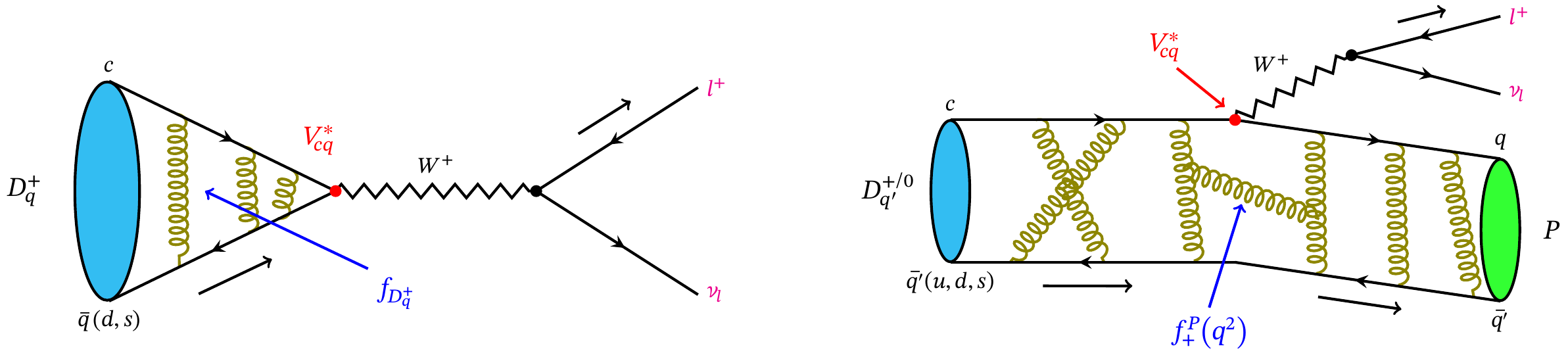}
  \caption{\small Diagrams for purely leptonic (left) and semi-leptonic (right) decays of $D_{(s)}$ mesons. (Courtesy of Hao-Kai Sun, Institute of High Energy Physics)}
\label{fig:dpl}
\end{figure}

Purely leptonic and semi-leptonic decays of hadrons have a special characteristic advantage in studies of  the weak interaction~\cite{Khlopov:1978id,Gershtein:1976mv}. 
A key feature is their
relative simplicity, a consequence of the fact that in these processes the effects of the strong interactions can be isolated. 
For each decay type, the decay amplitude can be written as the product
of a well-understood leptonic current for the process $W^+\to\ell^+\nu_\ell$ ($\ell$ denotes charged leptons)
and a more complicated hadronic current for the quark transition. Figure~\ref{fig:dpl} shows the Feynman diagrams for the purely leptonic (left  in Fig.~\ref{fig:dpl}) and  semi-leptonic decays (right in Fig.~\ref{fig:dpl}). 
In purely leptonic decays, the hadronic current describes an annihilation of the quark and the anti-quark in the initial-state
charmed mesons, while in semi-leptonic decays it describes the evolution from the initial-charmed  hadron to the final-state hadrons. Because strong
interactions affect only one of the two currents, purely leptonic and semi-leptonic decays are relatively simple from a theoretical
perspective; they provide bilateral means both to measure fundamental SM parameters and to perform detailed
studies of the decay dynamics~\cite{Richman:1995wm}.  

\subsection{A bridge between quark and lepton: decay constants and lepton flavor universality }

Purely leptonic decays of the $D^+$  and $D^+_s$  mesons are among the simplest and best-understood
probes of $c \to d$ and $c \to s$ quark transitions.
 In each case, the effects of the strong interaction can be parametrized in 
terms of just one factor, called the decay constant  $f_{D_{q}^+}$.  In SM, the corresponding decay rate, ignoring radiative corrections,
is given in a  simple form: 
\begin{equation}
\Gamma(D_{q}^+ \to \ell^+\nu_\ell)=
     \frac{G^2_F f^2_{D_{q}^+}} {8\pi}
      \mid V_{cq} \mid^2
      m^2_\ell m_{D_{q}^+}
    \left (1- \frac{m^2_\ell}{m^2_{D_{q}^+}}\right )^2,
\label{eq01}
\end{equation}
where  $q = d$ or $s$ quark and $ \ell= e$, $\mu$ or $\tau$ (electron, muon or tau lepton), and $\nu_\ell$ stands for the neutrino with the corresponding lepton flavor. 
 The $D^+_q$ mass ($m_{D_{q}^+}$), the mass of the charged lepton ($m_{\ell}$)  and the Fermi coupling constant ($G_F$) are all known to high precision~\cite{PDG}. 
Thus, the determination of $\Gamma(D_{q}^+ \to \ell^+\nu_\ell)$ directly measures the product
 $f_{D_{q}^+} |V_{cq}| $ of the $ D^+_q$ decay constant and the magnitude of the $c\to q$  CKM matrix element. 
One can then either extract $|V_{cq}|$ by using the predicted value of  $f_{D_{q}^+}$, $e.g.$, from LQCD~\cite{Bazavov:2018omf}, or obtain $f_{D_{q}^+}$ by using the experimentally measured $|V_{cq}|$ to test the LQCD prediction.  

Since the purely leptonic decays of pseudoscalar mesons are helicity-suppressed,   their decay
rates are proportional to the square of the charged lepton mass.  According to Eq.~\eqref{eq01}, the SM-expected
relative decay widths for the $\tau \nu_\tau$, $\mu \nu_\mu$ and $e\nu_e$ modes are $2.67:1:2.35\times 10^{-5}$ for $D^+$ and $9.75:1:2.35\times 10^{-5}$ for $D^+_s$ with negligible uncertainties. 
Therefore, the SM $D^+_q \to e^+\nu_e$ branching fractions  are expected to be $\BR_{e^+\nu_e} < 10^{-8}$ and  not yet
experimentally observable.

 \begin{figure}[hbp]
\centering
  \subfigure[]{\includegraphics[width=2.6in]{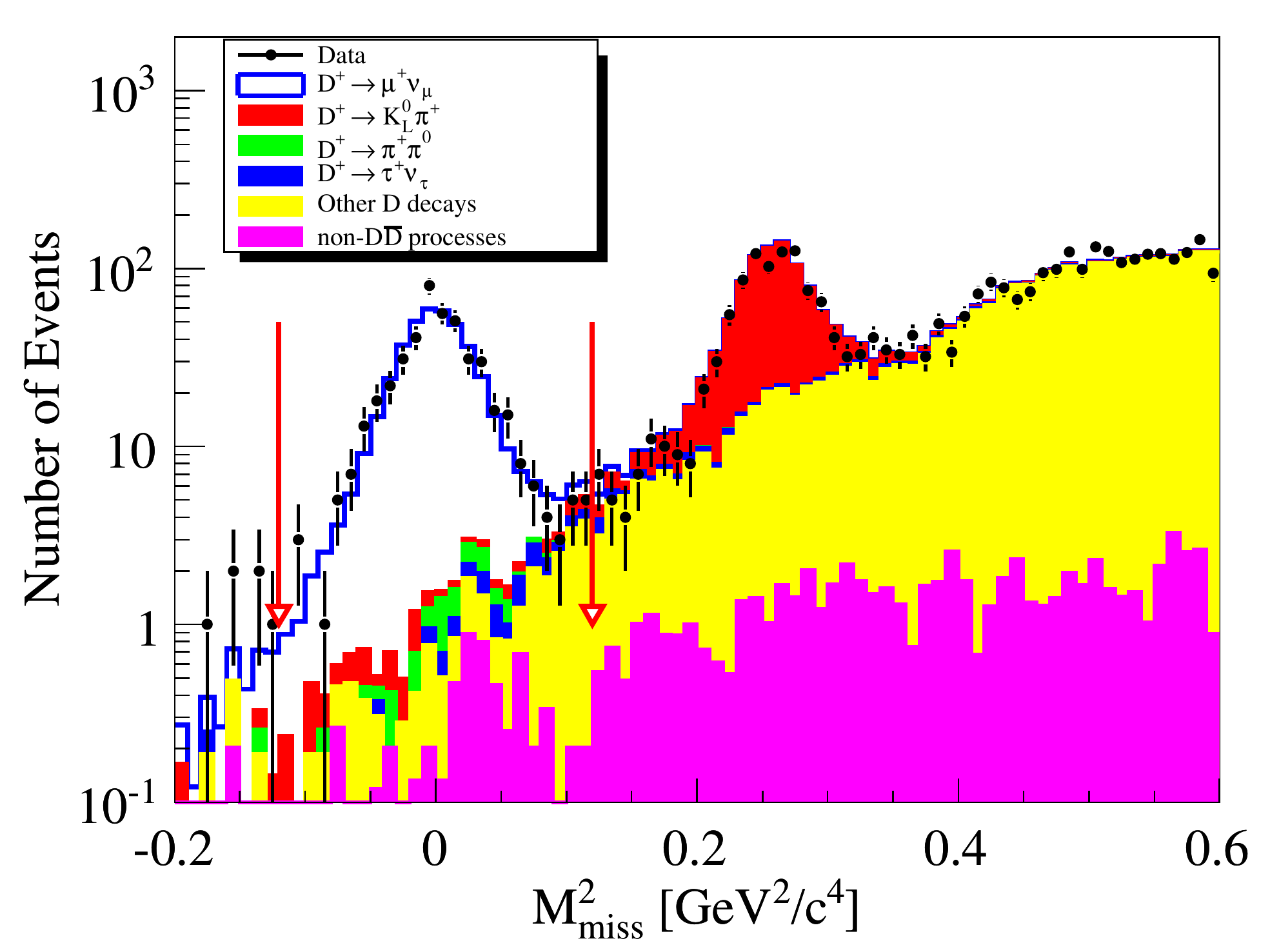}}
  \subfigure[]{\includegraphics[width=2.6in]{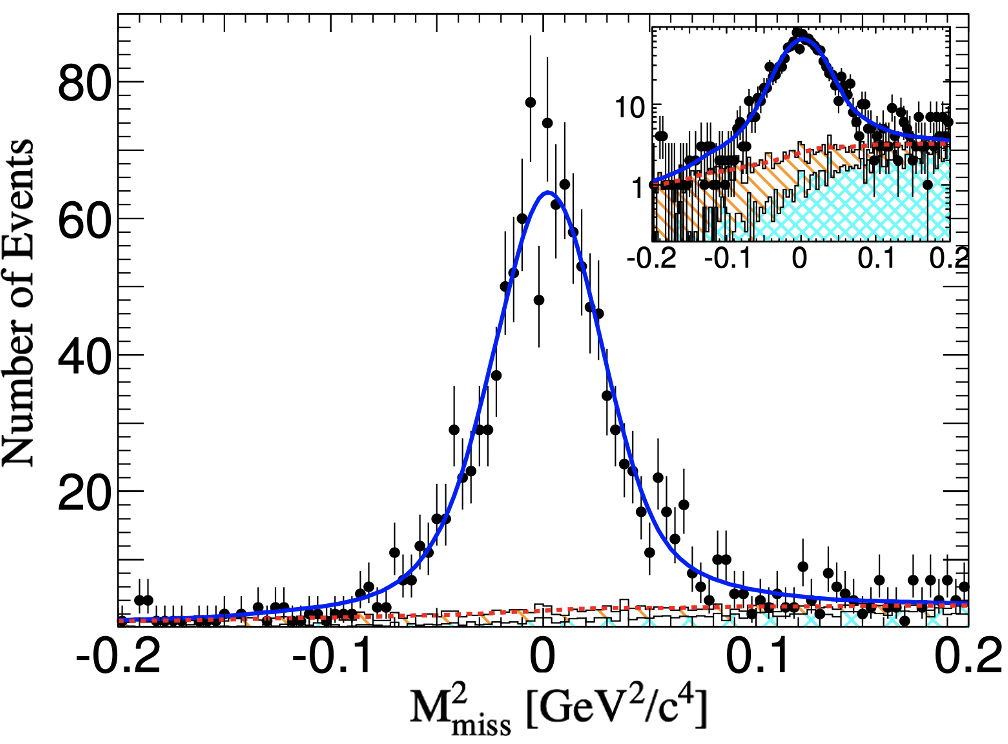}}
  \caption{\small The missing-mass $\mmiss^2$ distribution of the selected (a) $D^{+} \to \mu^{+} \nu_\mu$ and (b) $D^{+}_s \to \mu^{+} \nu_\mu$ candidates from Refs.~\cite{Ablikim:2013uvu} and~\cite{Ablikim:2018jun}, respectively.
   The error bars show the statistical uncertainty in experimental data. Arrows in plot (a) are the boundaries of the signal region, and the inset in plot (b) shows the same distribution in  logarithmic scale. Plots are from Refs.~\cite{Ablikim:2013uvu} and~\cite{Ablikim:2018jun}.}
\label{fig:dmunu}
\end{figure}

\begin{table}[htp]
\centering
\caption{\label{tab:pure}
 Measurements of $D^+$ and $D^+_s$ purely leptonic decays with threshold data at BESIII,  and comparisons between experimental results and theoretical expectation or SM-global fit results. (Here, ``-" 
 indicates not available.)}
\begin{footnotesize}
\begin{tabular}{lccc}
\hline\hline
Observable & Measurement  & Prediction/Fit \\ \hline
$\mathcal B(D^+\to \mu^+\nu_\mu)$& $(3.71 \pm 0.19_\mathrm{stat}\pm 0.06_\mathrm{syst}) \times 10^{-4}$~\cite{Ablikim:2013uvu} &- \\
$f_{D^+}|V_{cd}| $ & $(45.75 \pm 1.20_\mathrm{stat}\pm 0.39_\mathrm{syst})$ MeV&-  \\ 
$f_{D^+} $  &   $(203.8 \pm 5.2_\mathrm{stat} \pm 1.8_\mathrm{syst})$ MeV & $ (212.7 \pm 0.6)$ MeV~\cite{Bazavov:2018omf}\\
$|V_{cd}| $ &  $0.2150 \pm 0.0055_\mathrm{stat}  \pm 0.0020_\mathrm{syst}$ &  $0.22438 \pm 0.00044$~\cite{PDG} \\ \hline 
$\BR(D^+ \to \tau^+ \nu_\tau) $ &  $(1.20 \pm 0.24_\mathrm{stat}\pm 0.12_\mathrm{syst}) \times 10^{-4}$~\cite{Ablikim:2019rpl} &- \\  \hline 
$\Gamma(D^+ \to \tau^+ \nu_\tau)/  \Gamma(D^+ \to \mu^+ \nu_\mu)$ & $3.21 \pm 0.64_\mathrm{stat}\pm 0.43_\mathrm{syst}$~\cite{Ablikim:2019rpl} &  2.67 \\ \hline 
$\mathcal B(D^+_s \to \mu^+\nu_\mu)$& $(5.49 \pm 0.16_\mathrm{stat}\pm 0.15_\mathrm{syst})\times 10^{-3}$~\cite{Ablikim:2018jun} &-  \\
$f_{D^+_s}|V_{cs}| $ & $(246.2 \pm 3.6_\mathrm{stat}\pm 3.5_\mathrm{syst})$ MeV & -  \\ 
$f_{D^+_s} $  &   $(252.9  \pm 3.7_\mathrm{stat} \pm 3.6_\mathrm{syst})$ MeV & $ (249.9 \pm 0.5)$ MeV~\cite{Bazavov:2018omf, Carrasco:2014poa} \\
$|V_{cs}| $ &  $0.985 \pm 0.014_\mathrm{stat}  \pm 0.014_\mathrm{syst}$ &  $0.97359 \pm 0.00011$~\cite{PDG}\\ \hline 
$\Gamma(D^+_s \to \tau^+ \nu_\tau)/  \Gamma(D^+_s \to \mu^+ \nu_\mu)$ & $9.98\pm 0.52$~\cite{Ablikim:2018jun} & 9.74\\ \hline 
$f_{D^+_s}/f_{D^+}$  & $1.24 \pm   0.04_\mathrm{stat}\pm 0.02_\mathrm{syst}$~\cite{Ablikim:2018jun} & $1.1783 \pm 0.0016$~\cite{Aoki:2019cca}  \\
\hline\hline
\end{tabular}
\end{footnotesize}
\end{table}

Using a data sample with an integrated luminosity of $2.93$ fb$^{-1}$ collected with BESIII  at the $\psi(3770)$ peak,
a total of about 1.7 million single-tag $D^{-}$ mesons are selected using nine hadronic decay modes (summing up to 30\% of all $D^-$ decays) on the
tagging side. Throughout  this article,  charge-conjugate modes are implicitly assumed, unless otherwise stated.
Signal candidates of $D^{+} \to \mu^{+} \nu_\mu$ are required
to have a signature in which the tagging $D^{-}$ mesons are accompanied by exactly one track that is identified as a muon with charge opposite to that of the tagging $D^{-}$.  Since the massless neutrino is undetected,
 the yields of the signal  $D^{+} \to \mu^{+} \nu_\mu$ decays are measured based on the missing-mass-squared
variable $\mmiss^2 = (\ebeam - E_{\mu})^2 - (-\pvect  - \vec{p}_\mu )^2$, where $E_{\mu}$ and $\vec{p}_\mu$ are the energy and three-momentum of the muon, respectively, and $\pvect$ is the three-momentum of the
tagged $D^{-}$ candidate. 
$\mmiss$ corresponds to the invariant mass of the neutrino, and hence
the signal for $D^{+} \to \mu^{+} \nu_\mu$ events is the peak around $\mmiss^2 = 0$ as shown in Fig.~\ref{fig:dmunu} (a), where tiny background is smoothly distributed under the signal peak.  
From this,  BESIII obtained the world's most accurate branching fraction measurement for $D^+ \to \mu^+ \nu_\mu$ decay~\cite{Ablikim:2013uvu} as shown in Table~\ref{tab:pure}.   By inputting either 
the  LQCD-calculated value for the decay constant~\cite{Bazavov:2018omf} or the CKM matrix element values from a global SM fit~\cite{PDG} ,   the $|V_{cd}| $ or $f_{D^+}$ can be extracted; the corresponding results are listed in Table~\ref{tab:pure}. 
Using the same data sample, BESIII recently reported the first measurement of the absolute
decay branching fraction for $D^+ \to \tau^+ \nu_\tau$  with a significance of $5.1 \sigma$~\cite{Ablikim:2019rpl}.   The presence of additional final-state neutrinos
from the $\tau^+$ decays results in  more background and a relatively larger systematic uncertainty  than in the  $D^{+} \to \mu^{+} \nu_\mu$ decay measurement.

To study the  $D^+_s \to \mu^+ \nu_\mu$ signal channel,  BESIII uses $\EE \to D_s^+ D^{*-}_s$ collisions at the center-of-mass energy of $4178$ MeV and performs a similar analysis as was used for the $D^+ \to \mu^{+} \nu_\mu$ decay measurement~\cite{Ablikim:2018jun};  the $D^+_s \to \mu^+ \nu_\mu$ signal peak is   shown in Fig.~\ref{fig:dmunu} (b).   The absolute branching fraction and the product $f_{D^+_s}|V_{cs}| $ are obtained as listed in  Table~\ref{tab:pure}. Taking the CKM matrix element $|V_{cs}|$
 from the latest global SM fit~\cite{PDG},  the $D^+_s$ decay constant is determined.  Alternatively, taking the averaged decay constant 
 from recent LQCD calculations~\cite{Bazavov:2018omf, Carrasco:2014poa}, the CKM matrix element is extracted as listed in Table~\ref{tab:pure}.   These are the most precise measurements to date, and provide an  important  calibration of the theoretical calculations of $f_{D^+_s}$ and a stringent test of the unitarity of the CKM matrix with an improved  accuracy. 
 
   Using the world average values from the Particle Data Group (PDG)~\cite{PDG},  we determine the ratio 
   \begin{equation}
   R^{D^+}_{\tau/\mu} = \Gamma(D^+ \to \tau^+ \nu_\tau)/  \Gamma(D^+ \to \mu^+ \nu_\mu) = 3.21 \pm 0.64_\mathrm{stat}\pm 0.43_\mathrm{syst}, 
 \label{eq-pure1}
\end{equation}
which,  although  still statistically limited, is consistent with the SM prediction of 2.67 
With BESIII's expected future 20 fb$^{-1}$ data set at the $\psi(3770)$ peak, as discussed in Ref.~\cite{Ablikim:2019hff} and approved by the  collaboration, the precision on $R^{D^+}_{\tau/\mu} $ will be statistically improved to about 8\%,  which will provide an important test of the lepton flavor universality (LFU). 
 For the $D_s^+$,  we obtain 
   \begin{equation}
 R^{D^+_s}_{\tau/\mu} = \Gamma(D^+_s \to \tau^+ \nu_\tau)/  \Gamma(D^+_s \to \mu^+ \nu_\mu) = 9.98\pm 0.52, 
\label{eq-pure2}
\end{equation}
which agrees with the SM-predicted value of 9.74. 
 Meanwhile,  $D^+_s \to \tau^+ \nu_\tau$ decays are currently  being studied at BESIII with an expected result
that will have a precision comparable to that achieved for  the $D^+_s \to \mu^+ \nu_\mu$ decay mode.  This result should improve the accuracy of the $f_{D^+_s}|V_{cs}| $ measurement and can also be used to test LFU in the ratio  $R^{D^+_s}_{\tau/\mu}$  with a precision of 4.7\% based on the current data set~\cite{Ablikim:2019hff}.  With the expected 6 fb$^{-1}$ data set at 4178 MeV, as discussed in Ref.~\cite{Ablikim:2019hff}, the precision on $R^{D^+_s}_{\tau/\mu} $ will be systematically limited at about 3\% or less,  which will provide for the most stringent test of the $\mu$-$\tau$ LFU in heavy quark decays~\cite{Amhis:2019ckw}. 

 Finally, combining the measured  $f_{D^+_s}|V_{cs}| $ value with its $f_{D^+}|V_{cd}| $ counterpart,  along with the $|V_{cd}/V_{cs}|$ value from the global SM fit~\cite{PDG},  BESIII made a direct measurement of the   $f_{D^+_s}/f_{D^+}$ decay constant ratio~\cite{Ablikim:2018jun} that is 1.5$\sigma$ higher than  the Flavour Lattice Averaging Group (FLAG) world average value~\cite{Aoki:2019cca}, as shown in Table~\ref{tab:pure}.  Since  LQCD can make a very accurate prediction of $f_{D^+_s}/f_{D^+}$, which is a unique property  of  purely leptonic $D^+/D^+_s$ decays,   BESIII can make unambiguous measurements of fundamental SM parameters and perform detailed studies of the charmed hadron decay dynamics.  For these purposes,  more data are  needed at the $D\bar{D}$ and $D_s^+ D_s^{*-}$ thresholds to pursue  high-precision calibrations of LQCD  calculations~\cite{Ablikim:2019hff}.    
   
\subsection{Precision measurements of the transition form factors } 
\label{sec:exclu}

In the SM,  semi-leptonic decays of charmed hadrons  involve the interaction of a leptonic current with a hadronic
current.   The latter is non-perturbative and cannot be calculated from  the first principles; thus it is
usually parameterized in terms of form factors.  Still,  the weak and strong effects in semi-leptonic  decays can be well-separated, since there are no strong final-state interactions between the leptonic and hadronic systems.  Among the semi-leptonic decays, the simplest case is $D^{0/+} \to P \ell^+ \nu_\ell$ ( where $P$ denotes a pseudoscalar meson),  for which the differential partial decay width  is given,  in
the limit of negligible charged lepton mass,  by  
\begin{equation}
\frac{{\mathrm d} \Gamma(D^{0/+} \to P \ell^+ \nu_\ell )}{{\mathrm d}q^2}=
     \frac{G^2_F\mid V_{cq} \mid^2 }{24 \pi^3}  p^3_P \mid f^P_+(q^2)  \mid^2 .    
     \label{eq02}
\end{equation}
$p_P$ is the magnitude of the three-momentum of the $P$ meson in the $D^{0/+}$ rest frame and $f^P_+(q^2) $ is the form factor of the hadronic weak current depending on $q^2= |M(\ell^+\nu_\ell)|^2$, the square of the four-momentum transfer between the initial state $D^{0/+}$ and final-state $P$.  Thus,  semi-leptonic decays can be used to extract the product of a form factor normalization chosen to be at $q^2=0$ and a CKM matrix 
element:  $|V_{cq}| f^P_+ (0)$ . These decays allow for a robust determination of the $|V_{cs}|$ and $|V_{cd}|$ CKM matrix elements in conjunction with form factors determined from LQCD calculations. 
Alternatively,  by inputting CKM matrix elements one can determine the  form factors to provide high-precision tests of  LQCD calculations. 

The CLEO  experiment had made precision measurements of semi-leptonic charm-decay rates using a data set accumulated  at the $\psi(3770)$ peak~\cite{Besson:2009uv}.  With a three-times-larger data set,  BESIII reported improved measurements of the absolute decay rates and the form factors,  thereby assuming an important role in  the precision tests of LQCD calculations~\cite{Ablikim:2019hff}.

\begin{table}[htp]
\centering
\caption{\label{tab:semi}
  Measurements of $D^0 /D^+$, $D^+_s$ and $\Lambda_c$ semi-leptonic decays with near-threshold data at BESIII,  and comparisons between experimental results and theoretical expectations or SM-global fit result. (Here  ``-"  indicates not available.)} 
\begin{footnotesize}
\begin{tabular}{lccc}
\hline\hline
Observable & Measurement  & Prediction/Fit \\ \hline
$\BR(D^0 \to K^-  e^+ \nu_e) $ & $(3.505 \pm 0.014_\mathrm{stat}\pm 0.033_\mathrm{syst})\%$~\cite{Ablikim:2015ixa} &- \\
$|V_{cs}| f^K_+ (0) $ & $0.7172 \pm 0.0025_\mathrm{stat}\pm 0.0035_\mathrm{syst}$~\cite{Ablikim:2015ixa} & - \\ 
$f^K_+ (0) $ & $0.7368 \pm 0.0026_\mathrm{stat}\pm 0.0036_\mathrm{syst}$~\cite{Ablikim:2015ixa} & $0.747 \pm 0.011\pm 0.015$~\cite{Aoki:2019cca}\\ \hline
$\BR(D^0 \to \pi^-  e^+ \nu_e)$ &$(0.295 \pm 0.004_\mathrm{stat}\pm 0.003_\mathrm{syst})\%$~\cite{Ablikim:2015ixa}  &- \\ \
$|V_{cd}| f^\pi_+ (0) $ & $ 0.1435\pm 0.0018_\mathrm{stat}\pm 0.0009_\mathrm{syst}$~\cite{Ablikim:2015ixa}  & - \\
$f^\pi_+ (0) $ & $0.6372 \pm 0.0080_\mathrm{stat} \pm 0.0044_\mathrm{syst}$~\cite{Ablikim:2015ixa}  &  $0.66\pm 0.02 \pm 0.02$~\cite{Aoki:2019cca} \\ \hline
$\BR(D^+ \to \bar{K}^0 e^+\nu_e) $ & $ (8.60 \pm 0.06_\mathrm{stat} \pm 0.15_\mathrm{syst})\%$~\cite{Ablikim:2017lks} & - \\ 
$f^K_+ (0) $ & $  0.725 \pm 0.004_\mathrm{stat}\pm 0.012_\mathrm{syst}$~\cite{Ablikim:2017lks} & $0.747 \pm 0.011\pm 0.015$~\cite{Aoki:2019cca}\\ \hline 
$\BR(D^+ \to \pi^0  e^+\nu_e ) $ & $ (0.363\pm 0.008_\mathrm{stat} \pm 0.005_\mathrm{syst} )\%$~\cite{Ablikim:2017lks} & - \\ 
$f^\pi_+ (0)  $ & $ 0.622 \pm 0.012_\mathrm{stat} \pm 0.003_\mathrm{syst}$~\cite{Ablikim:2017lks}  & $0.66\pm 0.02 \pm 0.02$~\cite{Aoki:2019cca} \\ \hline
 $f^\pi_+ (0)/f^K_+ (0) $ & $0.865\pm 0.013$~\cite{Ablikim:2017lks}  & $0.84\pm 0.04$~\cite{Ball:2006yd} \\ \hline 
 $\BR(\Lambda^+_c \to \Lambda e^+ \nu_e )$ & $(3.63 \pm 0.38_\mathrm{stat} \pm 0.20_\mathrm{syst})\%$~\cite{Ablikim:2015prg} & - \\ 
 $\BR(\Lambda^+_c \to \Lambda \mu^+ \nu_\mu)$ &  $(3.49\pm 0.46_\mathrm{stat} \pm 0.27_\mathrm{syst})\%$~\cite{Ablikim:2016vqd}  & - \\ 
 $\BR(\Lambda^+_c \to  \Lambda \mu^+ \nu_\mu)/ \BR(\Lambda^+_c \to  \Lambda e^+ \nu_e) $ & $0.96\pm 0.16_\mathrm{stat} \pm 0.04_\mathrm{syst}$~\cite{Ablikim:2016vqd} & $\approx 1.0$ \\
\hline\hline
\end{tabular}
\end{footnotesize}
\end{table}

\begin{figure}[hbp]
\centering
  \subfigure[]{\includegraphics[width=4.0in]{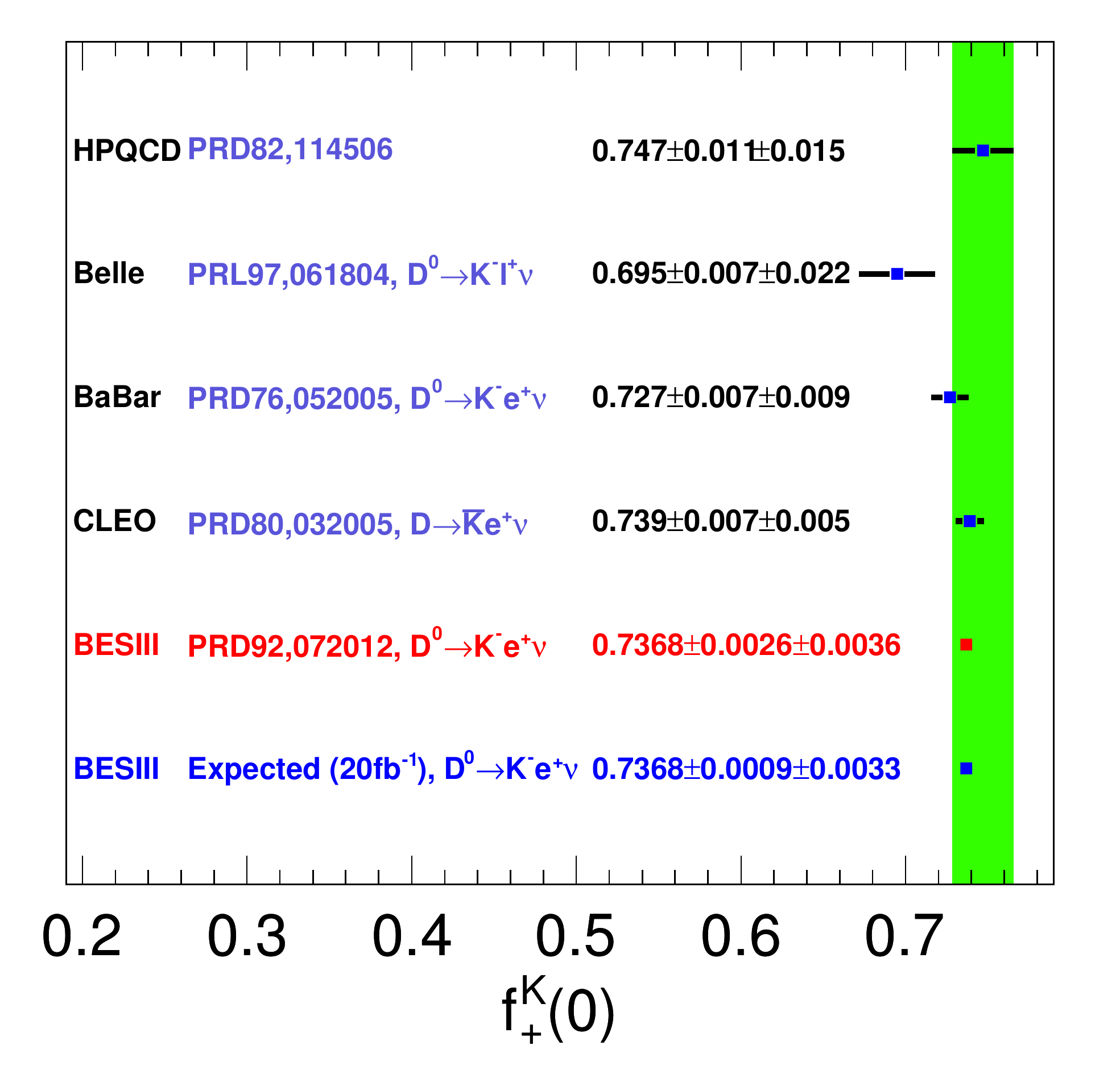}}
  \subfigure[]{\includegraphics[width=4.0in]{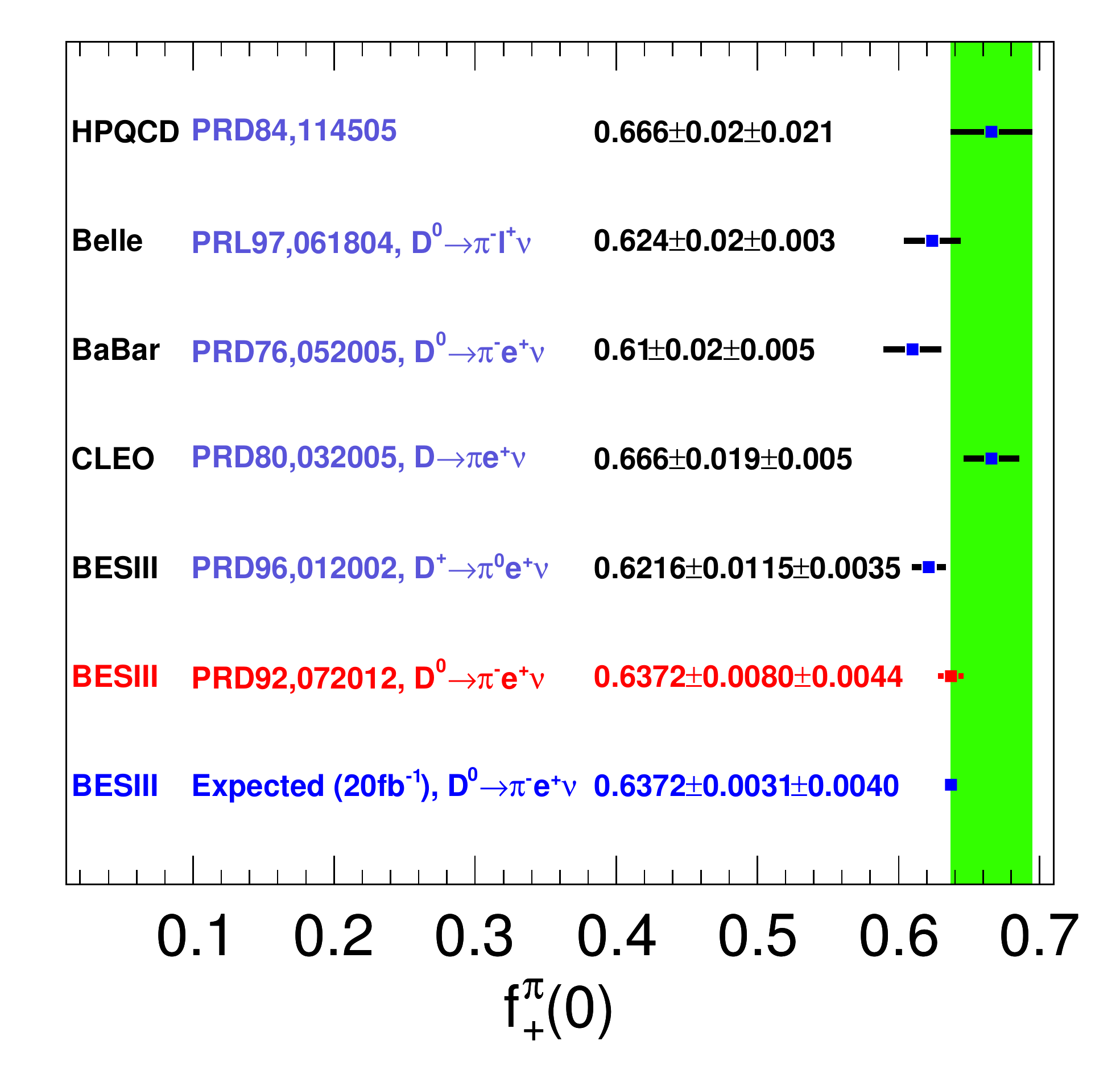}}
  \caption{\small Comparison of the results for (a) $f^K_+(0)$
and (b) $f^\pi_+(0)$  measured  by the Belle, BaBar, CLEO-c and BESIII experiments. The green bands present the LQCD uncertainties. 
The value marked in red denotes the best measurement from BESIII, and the value marked in dark blue denotes the
expected precision from BESIII with ten times the current data set~\cite{Ablikim:2019hff}. (Courtesy of Hai-Long Ma, Institute of High Energy Physics) }
\label{fig:ff}
\end{figure}

Notably, measurements  of the exclusive $D^0 \to K^- \ell^+\nu_\ell$ and $ \pi^- \ell^+\nu_\ell$ decays, as well as $D^+ \to \bar{K}^0 \ell^+\nu_\ell$ and $ \pi^0 \ell^+\nu_\ell$ decay modes, with $  \ell = e$ or $\mu$, have been reported~\cite{Ablikim:2015ixa, Ablikim:2017lks}.  Results for the absolute branching fractions are summarized  in Table~\ref{tab:semi}. 
 From studies  of the differential decay rates (see Eq.~(\ref{eq02})), 
the products of hadronic form factor at $q^2 =0$ and the magnitude of the CKM matrix element,  $|V_{cs}| f^K_+ (0)$  and $|V_{cd}| f^\pi_+ (0) $, are  shown in Table~\ref{tab:semi}.    Combining these products with the values of $|V_{cs}|$ and $|V_{cd}|$ from the SM-constrained fit~\cite{PDG}, we extract the transition form factors 
\begin{equation*}
 f^K_+ (0) = 0.7368 \pm 0.0026_\mathrm{stat}\pm 0.0036_\mathrm{syst}, 
     \label{eq-semi-1}
\end{equation*}
and 
\begin{equation*}
 f^\pi_+ (0)  =0.6372 \pm 0.0080_\mathrm{stat} \pm 0.0044_\mathrm{syst}, 
     \label{eq-semi-2}
\end{equation*}
and their ratio 
\begin{equation*}
\frac{f^\pi_+ (0)}{f^K_+ (0)} = 0.865\pm 0.013 , 
     \label{eq-semi-2}
\end{equation*}
which is in good agreement with present average ($0.834 \pm 0.023$) of LQCD calculations~\cite{Amhis:2019ckw,Aoki:2019cca} and 
a light cone sum rule value $0.84\pm 0.04$~\cite{Ball:2006yd}. 
 The experimental precision is  better  than that of  the theoretical calculation.  
The measurement of $f^\pi_+ (0)$ is dominated by statistical uncertainties.  More data will reduce these uncertainties as discussed in the BESIII future physics programme~\cite{Ablikim:2019hff}.  
Figure~\ref{fig:ff} shows the form factors $f^K_+ (0)$ and $f^\pi_+ (0)$ measured by various experiments together with results from LQCD calculations~\cite{Amhis:2019ckw,Aoki:2019cca}.

\begin{figure}[tp]
  \centering
     \includegraphics[width=2.5in]{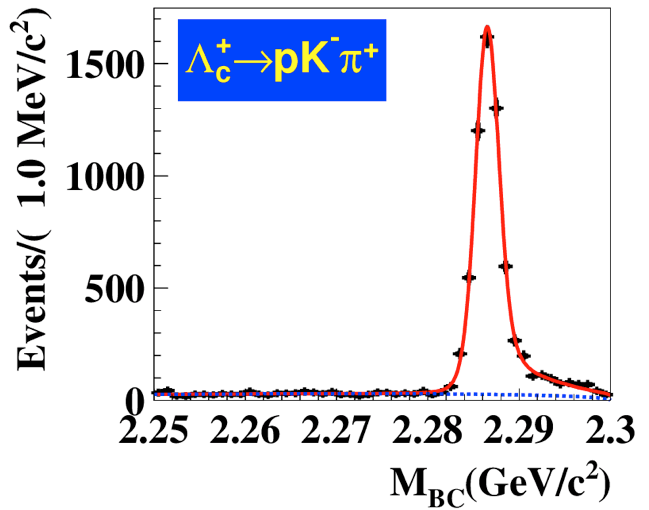}
    \caption{  Fit to the $M_{BC}$ distribution for $\bar{\Lambda}_c^- \to \bar{p} K^+ \pi^-$ decay in the tag side. 
The points with error bars are data, the solid
curves show the total fits, and the dashed curves are the background shapes. (Courtesy of Pei-Rong Li, Lanzhou University)}
\label{fig:Lc_signals-1}
\end{figure}

\begin{figure}[tp]
  \centering    
   \includegraphics[height=5.0cm,width=5.8cm]{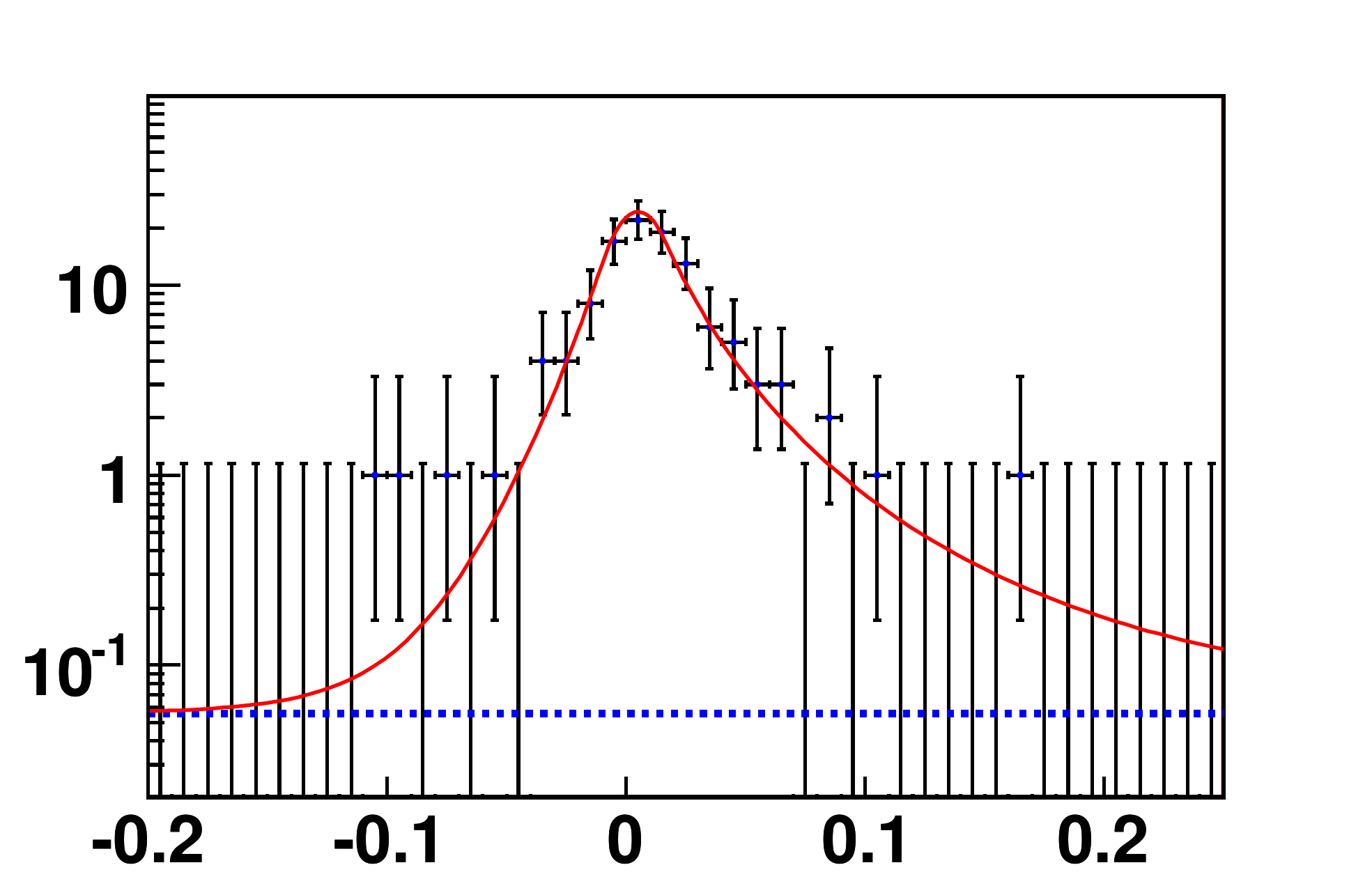}
   \put(-170,26){\rotatebox{90}{\footnotesize Events/0.010 GeV}}
   \put(-110,-10){\normalsize $U_{\rm miss}$ (GeV) }
    \caption{ Fit to the $U_{\rm miss}$ distribution within the $\Lambda$ signal region~\cite{Ablikim:2015prg}.
The points with error bars are data, the solid
curves show the total fits, and the dashed curves are the background shapes. Plot is from Ref.~\cite{Ablikim:2015prg}.}
\label{fig:Lc_signals-2}
\end{figure}

Based on a 567 pb$^{-1}$ data set collected at 4.6 GeV, an energy point slightly above the $\Lambda^+_c \bar{\Lambda}^-_c$ production  threshold,  BESIII made the first absolute branching fraction measurement of $\Lambda^+_c \to \Lambda e^+ \nu_e$~\cite{Ablikim:2015prg}.  Similar to the tagging technique employed in the $D\bar{D}$ threshold production, the  $\bar{\Lambda}_c^{-}$ is tagged via its hadronic decay modes.
As an example, Fig.~\ref{fig:Lc_signals-1} shows the beam-constrained mass $M_{\rm BC}$ distribution for the  $\bar{\Lambda}_c^- \to \bar{p} K^+ \pi^-$ tagging mode, where the background level is very low.  This 
is typical for most tagging modes and demonstrates that  the threshold data sets provide unique opportunities  for  nearly background-free charmed baryon decay measurements. 
Since the massless neutrino is undetected, 
the kinematic variable $U_{\text{miss}}= E_{\text{miss}}- c|\vec{p}_\text{miss}|$ is used  to infer its presence,  where $E_{\text{miss}}$ and $\vec{p}_\text{miss}$ are the missing energy and missing momentum carried by the neutrino, respectively. The calculation methods of  $E_{\text{miss}}$ and $\vec{p}_\text{miss}$ can be found in Ref.~\cite{Ablikim:2015prg}. The $U_{\text{miss}}$ distribution is presented in Fig.~\ref{fig:Lc_signals-2},  and a tiny background  under the signal peak is  inferred. 
 From this,  the absolute branching fractions for $\Lambda^+_c \to \Lambda e^+ \nu_e$ and $\Lambda^+_c \to \Lambda \mu^+ \nu_\mu$ decays are determined.  For the $\Lambda^+_c \to \Lambda e^+ \nu_e$ case, 
the BESIII result listed in Table~\ref{tab:semi} corresponds to a twofold improvement in the precision of the world average value.   Since the branching fraction for  $\Lambda^+_c \to \Lambda e^+ \nu_e$ is the benchmark and serves as a normalization mode for all other $\Lambda_c^+$ semi-leptonic channels, the BESIII result allows for stringent 
tests of different theoretical models. For the muonic decay $\Lambda^+_c \to \Lambda \mu^+ \nu_\mu$, the BESIII result  is the first direct measurement~\cite{Ablikim:2016vqd},  and with it the branching fraction ratio is determined to be: 
\begin{equation*}
\BR(\Lambda^+_c \to  \Lambda \mu^+ \nu_\mu)/ \BR(\Lambda^+_c \to  \Lambda e^+ \nu_e) = 0.96\pm 0.16_\mathrm{stat} \pm 0.04_\mathrm{syst}, 
     \label{eq-semi-2}
\end{equation*}
which is consistent with $e-\mu$ LFU.  The form factors for  charmed baryon transition to light hyperons/baryons will be studied with high precision when  more threshold data samples are collected by BESIII~\cite{Ablikim:2019hff}. 
The detailed $q^2$-dependent transition form-factors can be studied at BESIII, and will provide unique calibrations of LQCD calculations.  Alternatively, with LQCD predictions as input, BESIII measurements can be used to test LFU at any given $q^2$ value.    
  
 \begin{figure}[hbp]
\centering
  \subfigure[]{\includegraphics[width=3.8in]{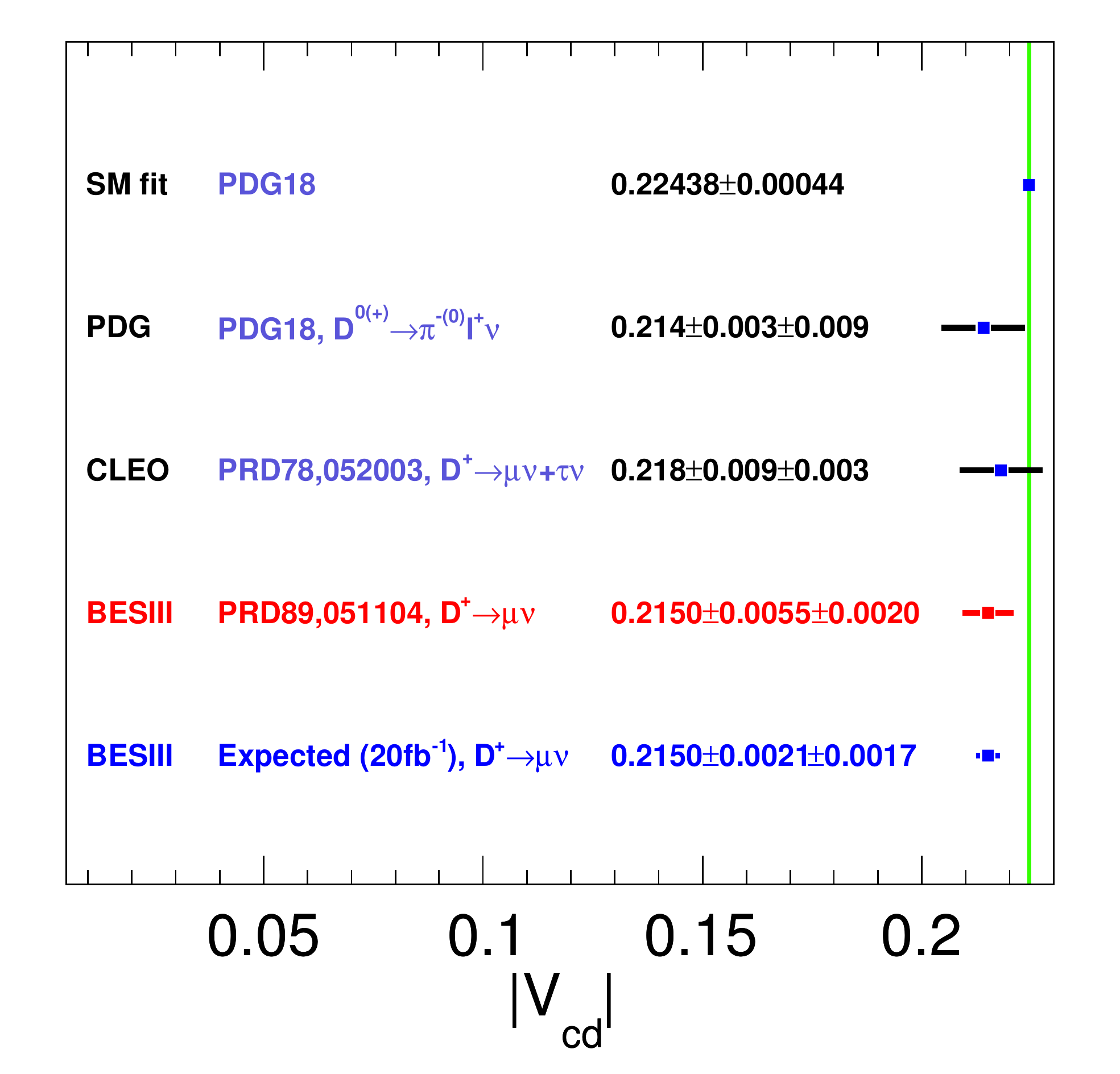}}
  \subfigure[]{\includegraphics[width=3.8in]{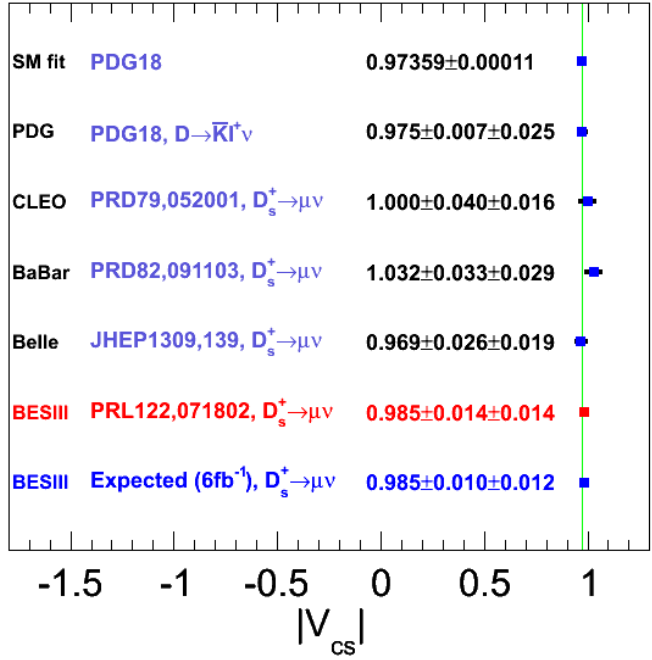}}
  \caption{\small  Precision of the measurements of: (a) $|V_{cd}|$; and (b) $|V_{cs}|$.   The green bands indicate the uncertainties of the average values from the global fit in SM~\cite{PDG2018}. The
circles, dots and rectangles with error bars are results derived from semi-leptonic $D$ decays, purely leptonic $D$ decays and other methods,
respectively. The results marked in red denotes the best
measurement, and the values marked in light blue denote the expected precision with the BESIII data sets that will be accumulated  in the future~\cite{Ablikim:2019hff}. (Courtesy of Hai-Long Ma, Institute of High Energy Physics)}
\label{fig:ckm}
\end{figure}

\begin{figure}[tp]
  \centering
    \includegraphics[width=3.2in]{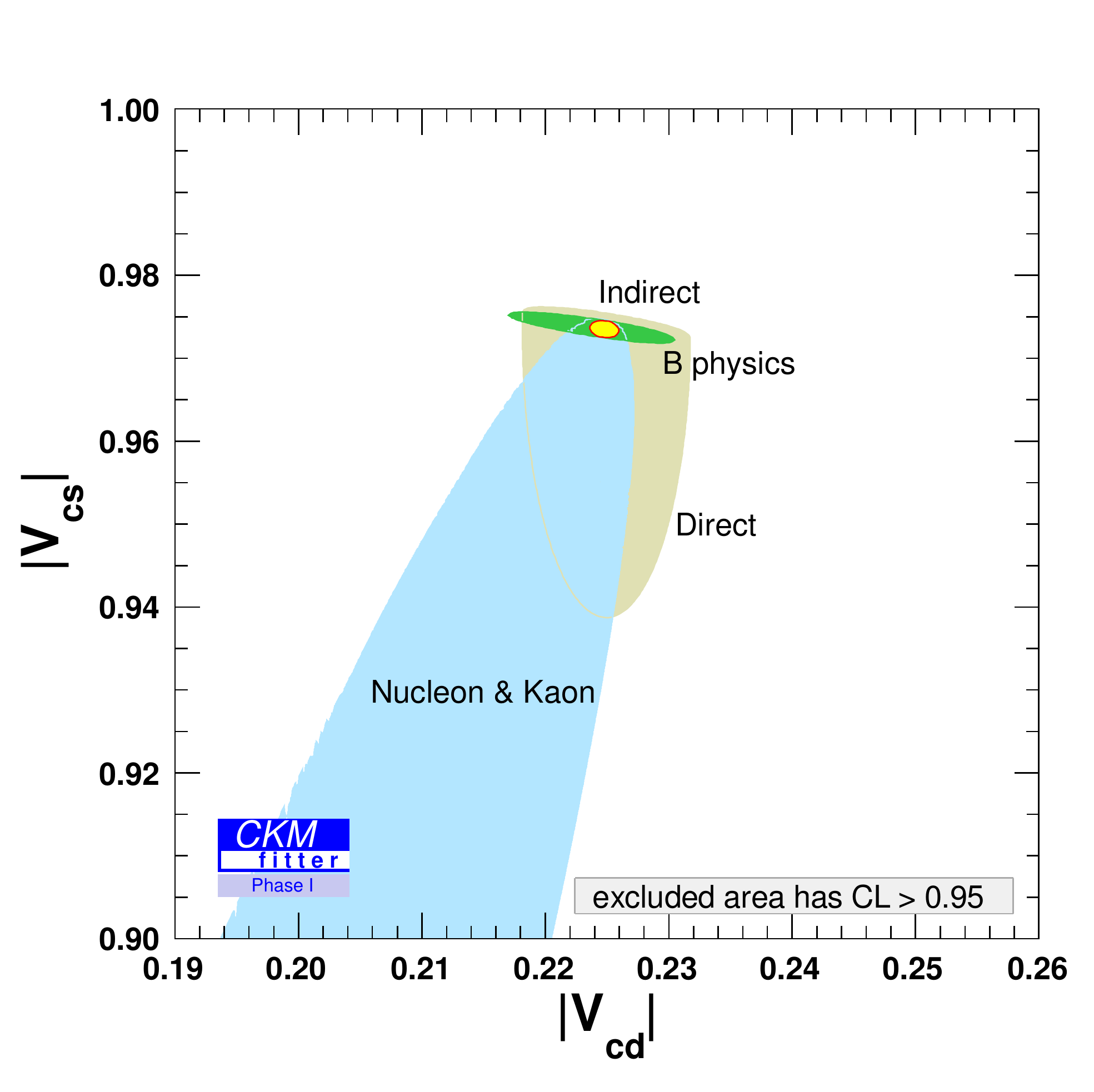}
   \caption{ Constraints on the  ($|V_{cs}|$, $|V_{cd}|$) plane expected  with future BESIII data-taking plan as described in the BESIII white paper~\cite{Ablikim:2019hff}. 
   The indirect (green) constraints (from $B$ decays) are related to $|V_{cs}|$ and $|V_{cd}|$ by unitarity~\cite{Charles:2004jd}. The direct (grey) constraints combine purely leptonic and semi-leptonic $D_{(s)}$ decays from
the BESIII experiment. The  red circled region of the global
combination corresponds to the 68\% confidence level. Plot is from ~\cite{Ablikim:2019hff}. }
\label{fig:vv_cs_cd}
\end{figure}

 \subsection{Impact on CKM matrix elements: $|V_{cs}|$ and$|V_{cd}|$ } 
 \label{sec:ckm}
 
  If precision LQCD calculations of the decay constants and form factors are taken as inputs,   measurements of branching fractions for  the purely leptonic and semi-leptonic decays can be  used to confront weak-interaction  physics.  In the past decade, great progress has been made in the LQCD calculations of decay constants.  The uncertainties of the results have been reduced from 
the level of (1-2)\% to  0.2\%~\cite{Aoki:2019cca}.  With these, the BESIII 
     leptonic-decay measurements have uncertainties of 2.5\% and 1.5\% for $|V_{cd}|$ and $|V_{cs}|$, respectively, and dominate the PDG world average values. For leptonic decays,  the statistical error on
    $|V_{cd}|$ is larger than the systematic error, while the statistical and systematic uncertainties of   $|V_{cs}|$ are comparable, as shown in Fig.~\ref{fig:ckm}.    The BESIII result for $|V_{cd}|$ listed in Fig.~\ref{fig:ckm} (a) is within 1.7$\sigma$ of the value obtained from a global SM fit to the other CKM matrix element measurements that assumes unitarity.  
    
     With additional data from the next ten year physics programme for BESIII~\cite{Ablikim:2019hff},   the relative errors on  the $|V_{cs}|$ and $|V_{cd}|$ determinations with purely leptonic decays will both reach the 1\% level; if the $|V_{cd}|$ result is the same as its current  central value,  the significance of the discrepancy would increase to about the 4 $\sigma$ level as shown in Fig.~\ref{fig:ckm} (a). 
     
      In addition,  with the FLAG~\cite{Aoki:2019cca} world average value for $(f_{D^+_s}/f_{D^+})^\mathrm{FLAG} = 1.1783 \pm 0.0016$,  BESIII obtained  $|V_{cd}/V_{cs}|^2 =  0.048\pm 0.003_\mathrm{stat}\pm 0.001_\mathrm{syst}$, which is consistent with the  one expected with the values of  $|V_{cs}|$ and $|V_{cd}|$  given by the CKMfitter group to within 2$\sigma$~\cite{Charles:2004jd}.   The error on the ratio $|V_{cd}/V_{cs}|^2$ is currently dominated by the limited  experimental statistics, and with the planned BESIII final data sample,  we expect  that the  statistical uncertainty  will be comparable to the systematic uncertainty which arises mainly from the LQCD decay constant calculations.  
     
The matrix elements $|V_{cs}|$ and $|V_{cd}|$ can also be determined from the measured partial widths for the semi-leptonic decays $D^{0(+)} \to \bar{K} \ell \nu_\ell$ and $D^{0(+)} \to \pi \ell \nu_\ell$ with 
     the computed values of the form factors from LQCD taken as inputs~\cite{Aoki:2019cca}.  The results using this method are also shown in Fig.~\ref{fig:ckm}. At present,  the uncertainties from LQCD calculations are 2.4\% for $f^K_+(0)$ and  4.4\% for $f^\pi_+(0)$, which are significantly larger than the uncertainties from the associated experimental measurements, and, therefore, limit the determinations of  $|V_{cs}|$ and $|V_{cd}|$ with this approach.  
     
      With the future BESIII data and improvements in the LQCD calculations on the decay constants and form factors that are expected circa 2025, we can anticipate significantly improved constraints on the ( $|V_{cs}|$, $|V_{cd}|$) plane as shown in Fig.~\ref{fig:vv_cs_cd}~\cite{Ablikim:2019hff},  where  direct contributions from the BESIII experiment are indicated. This will allow for precise tests of the consistency of CKM determinations from different quark sectors~\cite{Ablikim:2019hff, Charles:2004jd}.

\subsection{ Impact on charm lifetime and SU(3)$_{\rm F}$ symmetry from inclusive semi-leptonic decays}

Isospin symmetry requires that the  charged and neutral $D$ mesons
have the same inclusive semi-leptonic partial widths for Cabibbo-favored (CF) 
decays~\cite{Pais:1977nn}, and this is confirmed by experiments
within measurement uncertainties~\cite{PDG}.  This prediction 
is expected to be reliable, since the lepton cannot interact strongly with the final-state hadrons, and the  charged and neutral $D$ mesons
differ only in the isospin of the light quark.  Therefore,  a precise measurement of the  $\Gamma(D^0 \to X e^+ \nu_e)/\Gamma(D^+ \to X e^+ \nu_e)$ ratio ($X$ refers to any accessible hadronic system) provides a test of
 isospin symmetry.  With current values from the PDG~\cite{PDG},  one has  $\Gamma(D^0 \to X e^+ \nu_e) = (1.583 \pm 0.027 ) \times 10^{11}$ s$^{-1}$  and 
$\Gamma(D^+ \to X e^+ \nu_e) = (1.545 \pm 0.031) \times 10^{11}$ s$^{-1}$, and the observed ratio is 
 \begin{equation}
 R^{D^0/D^+}_{Xe\nu}=\frac{ \Gamma(D^0 \to X e^+ \nu_e)}{\Gamma(D^+ \to X e^+ \nu_e)} =1.025\pm 0.027.  
     \label{eq03}
\end{equation}
It indicates the need for reduction of experimental uncertainties
on the branching fraction measurements before the predicted deviations of this ratio
from unity can be identified.  On the other hand,  assuming the equality of semi-leptonic $D^0$ and $D^+$ partial widths, one obtains 
 \begin{align*}
 \tiny
 \frac{\tau_{D^+}}{\tau_{D^0}} = \frac{\Gamma(D^0 \to \mathrm{all})}{\Gamma(D^+ \to \mathrm{all} )} = \frac{ \Gamma(D^0 \to \mathrm{all})}{\Gamma(D^0 \to X e^+ \nu_e)} \frac{\Gamma(D^+ \to X e^+ \nu_e)} {\Gamma(D^+ \to \mathrm{all} )} 
  = \frac{\BR^{D^+}_{\rm SL}}{\BR^{D^0}_{\rm SL}}  ,   
     \label{eq0tau}
\end{align*}
where $\BR^{D^+}_{\rm SL}$ ($\BR^{D^0}_{\rm SL}$) is the inclusive semi-leptonic branching fraction for $D^+$ ($D^0$).  Therefore, comparison of  $\BR^{D^+}_{\rm SL}/ \BR^{D^0}_{\rm SL}$ with $\tau_{D^+}/\tau_{D^0}$  from direct lifetime measurements from other experiments provides a test of  isospin symmetry in charm decays and QCD calculations.  This analysis is currently ongoing at BESIII with triple the amount of CLEO-c data, and the sensitivity will be significantly improved.  

Furthermore,  inclusive semi-leptonic width measurements of
strange and non-strange $D$ mesons have revealed a clean determination  of SU(3) breaking effects. 
According to the operator product expansion methods~\cite{Gronau:2010if}, the partial widths for the inclusive semi-leptonic decays  of the $D^{+}$, $D^0$ and $D^+_s$ mesons
should be equal up to
SU(3)$_{\rm F}$  symmetry breaking and nonfactorizable contributions  (although their phase-space differences may not be
trivial~\cite{Gronau:2010if}). 
With the current values from the PDG~\cite{PDG},  one has  $\Gamma(D^+_s \to X e^+ \nu_e) = (1.300 \pm 0.082 ) \times 10^{11}$ s$^{-1}$, and the observed ratio is 
 \begin{equation}
R^{D_s/D}_{Xe\nu} = \frac{ \Gamma(D^+_s \to X e^+ \nu_e)}{\Gamma(D \to X e^+ \nu_e)} =0.830\pm 0.053. 
     \label{eq04}
\end{equation}
Thus, the semi-leptonic $D^+_s$ decay rate is $(17.0\pm 5.3)\%$  
lower than the charge-averaged non-strange $D$ semi-leptonic rate.
 This difference not only sheds light on strong-interaction dynamics, but can serve as a useful calibration
for measurements using $D^+_s$ decays~\cite{Gronau:2010if}.  
 Inclusive $D^+_s \to X e^+ \nu_e$ decay is
currently being studied at BESIII with an expected precision that will be comparable to that achieved for the corresponding  $D \to X e^+ \nu_e$ mode. 

 Similar to the cases for the charmed mesons ($D^0$/$D^+$/$D^+_s$) ,  the lifetime of the $\Lambda^+_c$ charmed baryon is dominated by the weak decay of the  charm quark, but  is is somewhat affected by the influence of the two accompanying light quarks ($u$ and $d$) in the hadron state in contrast to the single  light quark component in the meson case.  Therefore,  it will be interesting to make a comparison between the partial widths of the inclusive semi-leptonic decays  $\Lambda^+_c \to X e^+ \nu_e$ and   $D \to X e^+ \nu_e$, so that one can further understand the internal interactions and structures in the charmed baryon and mesons.  Information about exclusive semi-leptonic decays of the  $\Lambda^+_c$ 
 is sparse~\cite{PDG}, and only the $\Lambda^+_c \to \Lambda \ell^+ \nu_\ell$ ($\ell = e$ and $\mu$) decay mode has been measured. The measurement of the branching fraction of $\Lambda^+_c \to \Lambda \ell^+ \nu_\ell$ was first performed by the ARGUS collaboration~\cite{Albrecht:1991bu} and subsequently by the CLEO collaboration~\cite{Bergfeld:1994gt}.  Recently, the BESIII Collaboration measured the absolute branching fraction of  $\Lambda^+_c \to \Lambda \ell^+ \nu_\ell$ as discussed in Section~\ref{sec:exclu}~\cite{Ablikim:2015prg, Ablikim:2016vqd}.  A comparison of the exclusive semi-leptonic decay and  the inclusive semi-leptonic decay  will guide searches for new semi-leptonic decay modes.  Based on the threshold data at BESIII, the absolute branching fraction of the inclusive semi-leptonic decays of the $\Lambda^+_c$ baryon is determined to be 
 $\BR(\Lambda^+_c \to X e^+ \nu_e)= (3.95\pm 0.34_\mathrm{stat} \pm 0.09_\mathrm{syst})\%$~\cite{Ablikim:2018woi}, from which we obtain~\cite{Ablikim:2018woi} 
  \begin{equation*}
\BR(\Lambda^+_c \to  \Lambda e^+ \nu_e)/ \BR(\Lambda^+_c \to  X e^+ \nu_e) = (91.9\pm 12.5_\mathrm{stat} \pm 5.4_\mathrm{syst})\%, 
     \label{eq-inclusive-1}
\end{equation*}
 and  determine the ratio 
  \begin{equation*}
\Gamma(\Lambda^+_c \to  X e^+ \nu_e) / \Gamma(D\to X e^+ \nu_e) = 1.26\pm 0.12, 
     \label{eq-inclusive-1}
\end{equation*}
which can be used to restrict different QCD models and understand the internal interactions and structures in the charmed baryon and mesons~\cite{Gronau:2010if, Rosner:2012gj, Manohar:1993qn}.

 \section{ Unique probes  with quantum entangled $D^0\bar{D}^0$ and $\Lambda_c^+ \bar{\Lambda}_c^-$ states}

  BESIII operating at the $\psi(3770)$ resonance is a  ``charm factory''  that produces $D^0\bar{D}^0$  pairs  in a state of definite charge-conjugation eigenvalue $C=-$.  The antisymmetry of the wave function of the
   $D^0\bar{D}^0$   state induces
quantum entanglement  between the decay amplitudes of two $D$ mesons.  In particular, if one $D$ meson is reconstructed in a $CP$ eigenstate, the other $D$ meson is required to have the opposite $CP$ quantum number, provided $CP$ is  conserved in $D$ decays. 
Thus, the transition $\psi(3770) \to D^0\bar{D}^0$  occupies a special place in the charm experimentalist's and theorist's arsenal~\cite{Wilkinson:2021tby}.
  BESIII  data at $\psi(3770)$   offers crucial experimental advantages for the
determination of absolute branching fractions and interference between the two decay amplitudes from the entangled $D^0$ and $\bar{D}^0$ mesons~\cite{Ablikim:2014cea, Ablikim:2018fky,Ablikim:2015taz,Ablikim:2015wel},  that can be used to access their relative strong phases~\cite{Xing:1996pn}. This suite of measurements
is important to the international program in precision flavor physics and widely considered  to be one of the main motivations for a charm factory~\cite{Ablikim:2019hff}. Particularly pertinent to this review,
BESIII  offers unique opportunities to search for  $CP$ violation by exploiting quantum coherence in an almost  background-free environment.
   
   Analogously,  the reaction $e^+ e^- \to \Lambda^+_c \bar{\Lambda}^-_c$ produces charmed and anti-charmed baryon pairs.  The $\Lambda^+_c \bar{\Lambda}^-_c$ pair must be  in a $C$-odd quantum entangled state, which provides a unique opportunity to study the spin observables in the charmed baryon decays at  BESIII. 
  
  In this section, we discuss some selected measurements, including: 1) the unique quantum-coherent measurement of strong phases in neutral $D^0$ hadronic decays; and 2) the absolute branching fraction 
  measurements of the  $\Lambda_c^+$ hadronic decays firstly implemented  in the cleanest way near threshold. Both of these topics  are highlight results in charm physics at BESIII.

 \subsection{Relative strong phase and constraints on the  $CP$-violation phase $\gamma$}

\begin{figure}[tp]
  \centering
    \includegraphics[width=3.2in]{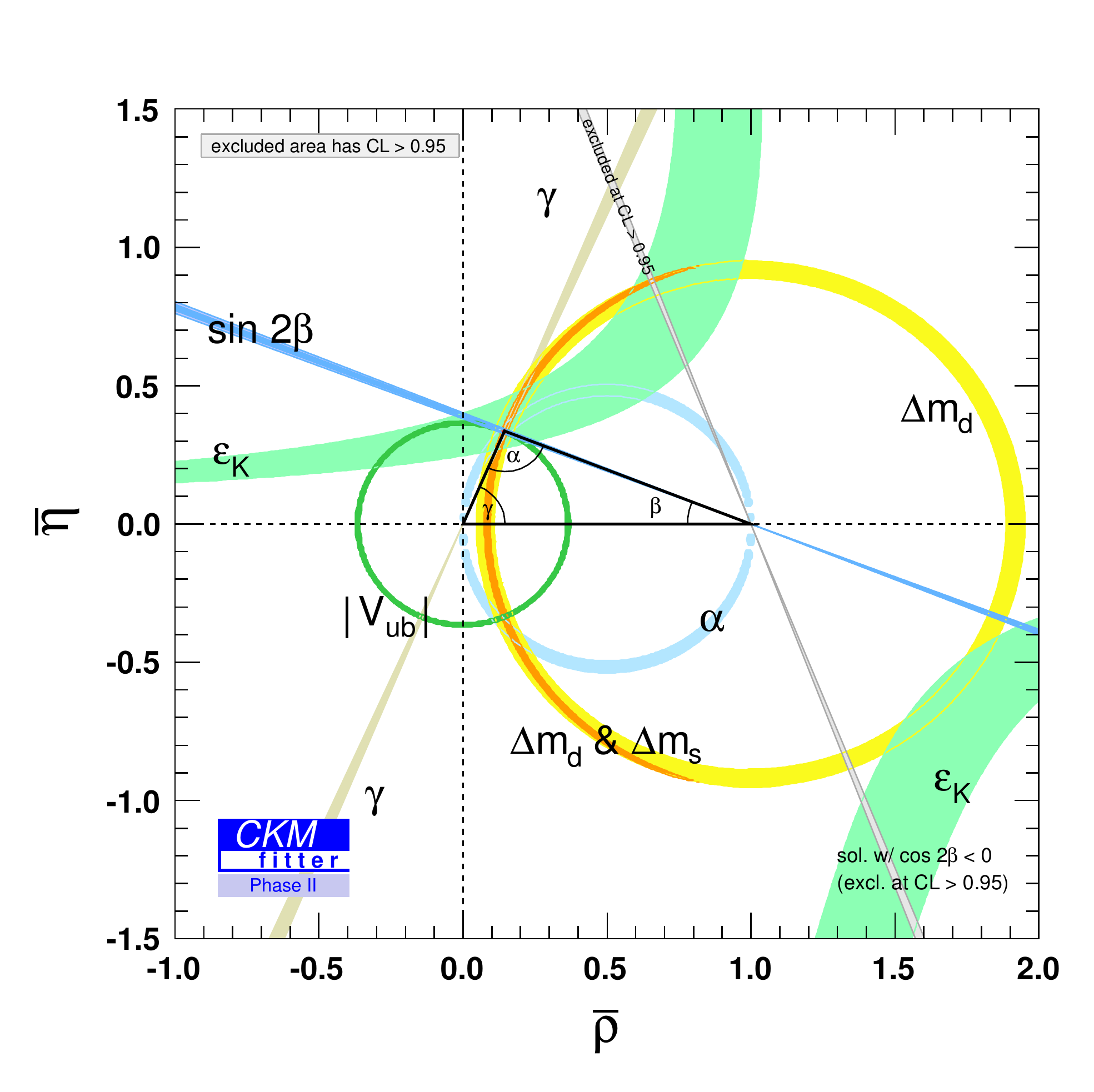}
   \caption{ Evolving constraints and the global fit in the $(\bar{\rho}, \bar{\eta})$ plane  (for $\bar{\rho}$ and $\bar{\eta}$ see  Eq. (\ref{eq:lambda:A:rhoeta})) with the anticipated improvements by considering the LHCb upgrade~\cite{Cerri:2018ypt} and final data set from the Belle II experiment~\cite{Kou:2018nap}.
 The $CP$-violation phase $\gamma$  is expected
to have an accuracy around 0.4 degrees, with the input from 20 fb$^{-1}$ of
data accumulated at the $D\bar{D}$ threshold~\cite{Ablikim:2019hff} at BESIII.  The shaded areas
have the 95\% confidence level.  The plot is from the CKM fitter group~\cite{Charles:2004jd} for BESIII future physics programme~\cite{Ablikim:2019hff}. Plot is from Ref.~\cite{Ablikim:2019hff}.}
\label{fig:vv_rho}
\end{figure}

The complex phases that  result from the strong interactions between the hadrons in the final states cannot be reliably calculated in theory and must be determined experimentally~\cite{Xing:1996pn}.  The values of the strong phase differences 
between the Cabibbo-favored (CF) and doubly Cabbibo-suppressed (DCS) amplitudes in charmed meson decays are crucial inputs for the extraction of the $CP$-violation phase angle $\gamma$,  {\it i.e.,} the phase of the CKM matrix element $V_{ub}$~\cite{Chau:1984fp,Wolfenstein:1983yz}, determined from measurements of $b$-hadron decays.  A precise measurement of $\gamma$ provides  a benchmark for tests of the SM that can be used as a probe to search for evidence of physics beyond SM~\cite{Charles:2004jd,Brod:2013sga,Asner:2008nq}.  
Three methods had been proposed to determine $\gamma$ so far:  GLW~\cite{Gronau:1990ra,Gronau:1991dp}, ADS~\cite{Atwood:1996ci,Atwood:2000ck}, and 
Dalitz (GGSZ)~\cite{Giri:2003ty} analyses.  One of the most sensitive decay modes  for measuring $\gamma$ is $B^- \to D K^- $ with $D \to K_S \pi^+\pi^-$~\cite{Giri:2003ty},   where 
 $D$ represents a superposition of $D^0$ and $\bar{D}^0$ mesons. The model-independent approach~\cite{Bondar:2005ki, Bondar:2008hh} requires a binned Dalitz plot analysis of the amplitude-weighted average cosine and sine of the relative strong-phase  between $D^0$ and $\bar{D}^0\to K_S \pi^+\pi^-$  decay amplitudes  to determine $\gamma$.
These relative strong phases can be uniquely determined from quantum-correlated $\psi(3770) \to D^0\bar{D}^0$ decays.

In 2009 and 2010, the CLEO experiment presented  first measurements of the strong-phase parameters by using 0.82\,fb$^{-1}$ of data~\cite{cleo-1,cleo-2}. The limited precision of these strong phase parameters translates into a  systematic uncertainty for the measurement of $\gamma$ of approximately 4 degrees~\cite{Aaij:2018uns}. In the coming decade, the statistical uncertainty of $\gamma$  is expected to be 1.5 degrees or less, in which case the overall precision will be limited by the strong-phase inputs. Hence, measurements of the relative strong-phase parameters with improved precision are a high priority activity that is  critical for a range of $CP$-phase measurements. 
At present,  BESIII is the only running experiment that can take data at the  $D^0\bar{D}^0$-pair-production threshold.  Based on 2.93\,fb$^{-1}$ of data, BESIII recently explored several methods to improve the analysis by incorporating more hadronic $D^0$ decays to increase statistics, including the development of partial reconstruction techniques to improve signal efficiency, and taking the effects of bin migration into account to reduce possible deviations of the results~\cite{Ablikim:2020lpk,Ablikim:2020yif,Ablikim:2020cfp}. These improvements are critical to provide better precision and accuracy compared to previous measurements~\cite{Aaij:2018uns}. Using the new BESIII results, the effect of the strong-phase uncertainty on the value of $\gamma$ will be reduced to around 1.0 degrees, which is approximately a factor of three smaller than what was possible with the CLEO measurements~\cite{cleo-1,cleo-2}. This will ensure that the anticipated statistical improvements in the measurements of $CP$-phase $\gamma$ at the LHCb and Belle II experiments over the next decade will  have  a deep impact on precision tests of the SM.

  Given the future sensitivities resulting from the LHCb upgrade~\cite{Cerri:2018ypt} and the final data set from Belle II~\cite{Kou:2018nap},  
  the overall constraints and the global CKM fit on the $(\bar{\rho}, \bar{\eta})$ plane (for $\bar{\rho}$ and $\bar{\eta}$ see Eq.~\eqref{eq:lambda:A:rhoeta}) are shown in Fig.~\ref{fig:vv_rho}. The $CP$-phase $\gamma$  is expected
to have an accuracy of around 0.4 degrees~\cite{Bediaga:2018lhg}, thanks to the BESIII charm input that will be available  with the full  data sample of 20\,fb$^{-1}$ at the $D\bar{D}$ mass threshold that is part of the experiment's future running plan~\cite{Ablikim:2019hff}. 
Based on the future largest $D\bar{D}$ sample, quantum-coherence measurements of strong phases of more charm decay modes, as stated in Ref.~\cite{Ablikim:2019hff}, facilitate stringent cross check of independent approaches of determining the $\gamma$ angle and provide constraints in world-wide averaging the $D^0$-$\bar{D}^0$ mixing parameters and the involved indirect CP violation.  The future high-statistics $B$ decay data at future LHCb upgrade provides sensitivity in accessing the strong phase parameters in principle. However, the final BESIII measurement is necessary input for reaching the target $\gamma$ sensitivity.

\subsection{Absolute branching fraction measurements of the  $\Lambda_c$ decays }  

Measurements of weak decays of charmed baryons provide useful information for understanding the interplay of weak and strong interactions, and are complementary to the information obtained from charmed mesons. 
The lightest charmed baryon $\lambdacp$, with quark configuration $udc$, serves as the cornerstone of  charmed baryon spectroscopy. 
However, the progress of the theoretical understanding of $\lambdacp$ decays has been slow~\cite{1992JGKorner, 1994TUppal, 1994PZenczykowsky, 1996LLChau,1997KKSharma, 1998YKohara, Ivanov98, Asner:2008nq}, mostly due to limited understanding of the non-perturbative effects in QCD theory in the  charmed baryon sector.

Before 2014, most  $\lambdacp$ decay branching fractions were obtained by measuring their ratios to the reference mode  $\lambdacp \rightarrow p K^-\pi^+$, thus introducing strong correlations and compounding uncertainties. 
The old, experimentally averaged branching fraction, $\BR({\lambdacp \rightarrow p K^-\pi^+})=(5.0\pm1.3)\%$, had a large uncertainty due to the introduction of model assumptions on $\lambdacp$ inclusive decays in these measurements~\cite{Jaffe:2000nw}. Furthermore, only about 40\% of the total decay rate had been measured and many modes were not identified, such as those with final-state neutrons.  
Therefore, comprehensive experimental measurements of various $\lambdacp$ hadronic decays play an important role in improving different theoretical calculations~\cite{Cheng:2015iom} and developing the QCD methodology in handling non-perturbative effects.

\begin{table}[tp]
  \begin{center}
  \footnotesize
\caption{Measurements of the $\Lambda_c^+$ hadronic decays (two-body CF,   neutron-involved and SCS decays) at BESIII, and their comparisons to the previous  world averages. For BESIII results, the first uncertainties are statistical and the second are systematic.}
\newcommand{\tabincell}[2]{\begin{tabular}{@{}#1@{}}#2\end{tabular}} \begin{tabular}{l|l|l} \hline\hline
Decay channel   &   BESIII (\%) & Previous world averages (\%)~\cite{pdg_2014}\\

{\bf Two-body~CF} & \\
  $pK^0_{S}$         &$1.52\pm0.08\pm0.03$~\cite{plb_bfs}    & $1.15\pm0.30$\\
 $\Lambda\pi^+$      &$1.24\pm0.07\pm0.03$~\cite{plb_bfs}   &$1.07\pm0.28$\\
  $\Sigma^0\pi^+$     &$1.27\pm0.08\pm0.03$~\cite{plb_bfs}   &$1.05\pm0.28$\\
  $\Sigma^+\pi^0$     &$1.18\pm0.10\pm0.03$~\cite{plb_bfs}     &$1.00\pm0.34$\\
 $\Sigma^+\omega$    &$1.56\pm0.20\pm0.07$~\cite{plb_bfs}    &$2.7\pm1.0$\\
 $\Xi^0 K^+$          &$0.590\pm0.086\pm0.039$~\cite{Lc_XiK}     &$0.50\pm0.12$\\
  $\Xi(1530)^0 K^+$       &$0.502\pm0.099\pm0.031$~\cite{Lc_XiK}    &$0.4\pm0.1$\\
$\Sigma^+\eta$        &$0.41\pm0.19\pm0.05$~\cite{Ablikim:2018czr}    &$0.70\pm0.23$\\
 $\Sigma^+\eta'$       &$1.34\pm0.53\pm0.19$~\cite{Ablikim:2018czr}    & First evidence \\
 $\Sigma(1385)^+\eta$  &$0.91\pm0.18\pm0.09$~\cite{Ablikim:2018byv}     &$1.22\pm0.37$\\
\hline

\hline
{\bf Neutron-involved} &   \\
 $nK^0_{S}\pi^+$          &$1.82\pm0.23\pm0.11$~\cite{Ablikim:2016mcr}   & First observation\\
 $\Sigma^-\pi^+\pi^+ $     &$1.81\pm0.17\pm0.09$~\cite{Ablikim:2017iqd}   &$2.1\pm0.4$\\
  $\Sigma^-\pi^+\pi^+\pi^0$ &$2.11\pm0.33\pm0.14$~\cite{Ablikim:2017iqd}    & First observation\\

\hline
{\bf SCS} &   \\
 $p\phi$              &$0.106\pm0.019\pm0.014$~\cite{plb_ppipi}     &$0.082\pm0.027$\\
  $p\eta$        &$0.124\pm0.028\pm0.010$~\cite{Ablikim:2017ors}    & First evidence\\
  $p\pi^0$       &$<0.027~\rm{at~90\%~C.L.}$~\cite{Ablikim:2017ors}    & First measurement \\
$p\pi^+\pi^-$        &$0.391\pm0.028\pm0.039$~\cite{plb_ppipi}    &$0.35\pm0.2$\\
 $pK^+K^-$ (non-$\phi$) &$0.0547\pm0.0130\pm0.0074$~\cite{plb_ppipi}    &$0.035\pm0.017$\\
 
\hline\hline
\end{tabular}
\label{tab:Lc_decay}
\end{center}
\end{table}

Based on a 567 pb$^{-1}$ data sample  accumulated at  4.6 GeV, BESIII has systematically investigated the production and
decays of the $\Lambda_c^+$~\cite{Ablikim:2019hff} for the first time using near-threshold data, which guarantee clean background and controllable systematics.
 BESIII provided absolute measurement of $\br{(\lambdacp \rightarrow p K^-\pi^+)}$ by counting the relative yields of the detected $\Lambda_c^+\bar{\Lambda}{}_c^-$ pairs over the single $\Lambda_c^+$,  with the result ($5.84\pm0.27_{\rm stat}\pm0.23_{\rm syst}$)\%~\cite{plb_bfs}. This has competitive precision to the result ($6.84\pm0.24^{+0.21}_{-0.27}$)\% reported by Belle~\cite{belle_pkpi} at nearly the same time and the combined precision of the two measurements  is 5.2\%, a five-fold reduction of the previous uncertainty~\cite{pdg_2014}.  Since  this mode is the golden channel for detecting  $\Lambda_c^+$ baryons in hadron collider experiments, the BESIII result impacts many aspects of  heavy flavor physics. For instance,  since the  $\Lambda_b^0$ decays primarily to $\lambdacp$~\cite{Dytman:2002yd,Rosner:2012gj},  it constrains  the measurement of $|V_{ub}|$ via $\Lambda_b^0\to \lambdacp \mu^- \nu$. Improved measurements of   $\lambdacp$ hadronic decays can be used to constrain charm and bottom quark fragmentation functions by counting inclusive heavy flavor baryons~\cite{Abreu:1999vw}. 

In addition, BESIII reported numerous absolute branching fraction measurements of two-body CF and singly Cabibbo-suppressed (SCS) decays  with improved precision, as listed in Table~\ref{tab:Lc_decay}.
The calculated  branching fractions for these  modes still have large uncertainties, and precise experimental measurements are essential to calibrate different models and explore the dynamics in charmed baryon decays. For instance, the improved  precision provides crucial input to the theoretical predictions~\cite{Yu:2017zst} for the observation channels of the doubly charmed baryon $\Xi_{cc}^{++}$ at LHCb~\cite{Aaij:2017ueg}. 
In particular, the improved precision of the SCS modes is useful for testing SU(3)$_{\rm F}$ symmetry in the charm sector and provides insight of the size of $CP$ violation  in the charmed baryon sector~\cite{Geng:2017mxn}. 

Moreover, BESIII observed, for the first time,  decay modes with a neutron in the final state, including $\Lambda_c^+\to nK^0_{S}\pi^+$~\cite{Ablikim:2016mcr} and $\Sigma^-\pi^+\pi^+\pi^0$ with $\Sigma^- \to n \pi^-$~\cite{Ablikim:2017iqd}. These analyses were carried out by using the missing-mass technique to infer the presence of a final-state  neutron that is only possible because of the kinematic constrains of  pair production in near-threshold data at BESIII. The results provide useful input to tests of  isospin symmetry in the charm sector.


\begin{figure}[tp]
  \centering
    \includegraphics[width=4.0in]{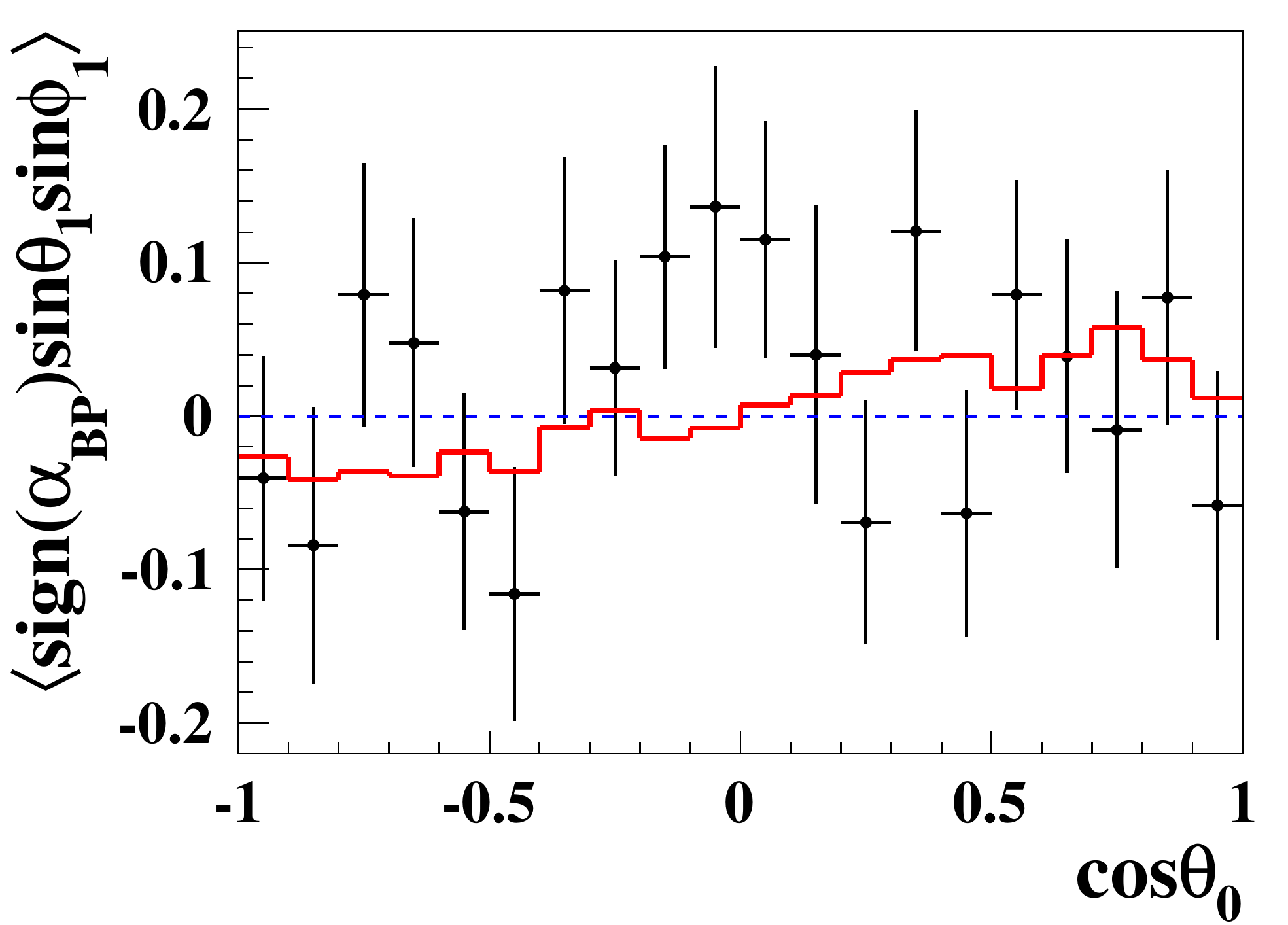}
   \caption{ The effect of the $\Lambda_c^+$ transverse polarization vs cos$\theta_0$  in $e^+e^- \to \Lambda_c \bar{\Lambda}_c$ at center-of-mass energy of 4.6 GeV.   Here $\theta_0$ 
   is the $\Lambda_c^+$  production angle relative to the $e^-$-beam direction. 
   The solid curve is fit to data; the dotted line is expectation for zero polarization. Detailed description can be found in Ref.~\cite{Ablikim:2019zwe}. Plot is from Ref.~\cite{Ablikim:2019hff}. }
\label{fig:lambda-c-p}
\end{figure}

The hadronic weak decays of charmed baryons are expected to violate parity conservation. For instance, in a two-body decay  $\Lambda_c^+\to BP$ ($B$ denotes a $J^P =\frac{1}{2}^+$ baryon and $P$ denotes a $J^P =0^-$ pseudoscalar meson) the parity asymmetry is defined  as
$\alpha^+_{BP}\equiv\frac{2{\rm Re}(s^* p)}{|s|^2+|p|^2}$, where
$s$ and $p$ stand for the parity-violating $s$-wave and
parity-conserving $p$-wave amplitudes in the decay, respectively.
For the process $\Lambda_c^+\to\Lambda\pi^+$, which proceeds via a $W$-interaction, $c\to W^+ + s$,  
the effects of parity violation are mainly determined by studying the polarization of the produced $\Lambda$ via its decays to $p\pi^-$ from the initially (polarized) charmed baryons~\cite{Wang:2016elx,Cheng:2015iom}.
If $CP$ is violated, the decay asymmetry parameters $\alpha^+_{BP}$  for $\lambdacp$ and $\bar{\alpha}^-_{\bar{B}\bar{P}}$ for $\lambdacm$ have different  magnitudes but are opposite in sign. Hence, separate determinations of $\alpha^+_{BP}$ and $\bar{\alpha}^-_{\bar{B}\bar{P}}$ would facilitate searching for the effects of $CP$ violation.
So far, only a few decay asymmetry parameters, $\egeg$, $\alpha_{\Lambda\pi}$ for $\Lambda_c^+\to\Lambda\pi^+$,  and $\alpha_{\Sigma^+\pi^0}$ for $\Lambda_c^+\to\Sigma^+\pi^0$, have been studied, and even those with limited precision~\cite{PDG}.
Therefore, improved or new decay asymmetry measurements are desirable, as they could shed light on the decay mechanism and allow searches for
$CP$ asymmetries in the charmed baryon sector. In addition the decay asymmetry values allow for discrimination between different theoretical models, as listed in Refs.~\cite{Cheng:2015iom,Cheng:2021qpd}.  

In the near-threshold production of  $\Lambda_c^+\bar{\Lambda}_c^-$ pairs,  non-zero transverse polarization of the $\Lambda_c^+$ 
will aid the determinations of the decay asymmetries.
The decay asymmetry parameters are determined by
analyzing the multi-dimensional angular distributions,
where the full cascade decay chains are considered. The detailed method can be found in Ref.~\cite{Ablikim:2019zwe}, in which 
a joint extraction of the four decay parameters of $\alpha_{\Lambda\pi}$, $\alpha_{\Sigma^+\pi^0}$, $\alpha_{\Sigma^0\pi^+}$ and $\alpha_{p\bar{K}^0}$ at the same time  was carried out based on the $\lambdacp$ sample at 4.6 GeV. An indiction of a non-zero transverse polarization is seen with a significance of 2.1$\sigma$ as shown in Fig~\ref{fig:lambda-c-p}, which makes the measurement of the
asymmetry parameter $\alpha_{p\bar{K}^0}$ accessible
experimentally for the first time.
The asymmetry parameters~\cite{Ablikim:2019zwe} for the $p\bar{K}^0$, $\Lambda\pi^+$, $\Sigma^+\pi^0$ and
  $\Sigma^0\pi^+$ modes are measured to be $
  0.18\pm0.43_{\rm stat}\pm0.14_{\rm syst}$,
  $-0.80\pm0.11_{\rm stat}\pm0.02_{\rm syst}$,
  $-0.57\pm0.10_{\rm stat}\pm0.07_{\rm syst}$, and
  $-0.73\pm0.17_{\rm stat}\pm0.07_{\rm syst}$, respectively.  In
  comparison with previous results, the measurements for the
  $\Lambda\pi^+$ and $\Sigma^+\pi^0$ modes are consistent but have an improved precision, while the parameters for the $p\bar{K}^0$ and
  $\Sigma^0\pi^+$ modes are measured for the first time. At present, no theoretical model provides predictions that are fully consistent with all these
measurements and BESIII measurements have become benchmarks to calibrate the QCD-derived theoretical models. During BESIII's 2020 data-taking,  about 10 times larger  $\Lambda_c^{+}$ samples were accumulated at the center-of-mass energies between 4.6 and 4.7 GeV, and  the significance of $\Lambda_c^+$ polarization could improve to  more than 5$\sigma$. In this case the precision of the decay asymmetries will be improved at least by a factor of 3.  With this information,  tests of $CP$ violations can be pursued for the two-body decays 
by comparing decay asymmetry parameters measured separately for $\Lambda_c^{+}$ and $\bar{\Lambda}_c^{-}$.

\section{ Summary and prospects } 
Charm particle weak  decays remain an exciting field for both theoretical
and experimental investigations. 
 In this article, we summarize results on  charm decays that have been obtained in the BESIII experiment with data sets collected at the production thresholds of $D\bar{D}$, $D_s^{*+} D^{-}_s$ and $\Lambda^+_c \bar{\Lambda}^-_c$.  These data samples allow the application of  double tag methods to fully reconstruct events
  even when invisible particles, such as neutrons or neutrinos, are present in the final states. This provides a unique environment to obtain the absolute branching fractions of charmed hadron decays to purely leptonic, semi-leptonic and hadronic final states with very low background levels. These BESIII measurements provide rigorous tests of QCD-based models and measurements of the CKM matrix elements $|V_{cs}|$ and $|V_{cd}|$, supply inputs to CKM weak phase measurements, and test leptonic-flavor universality. 
 
 Charmed hadron studies will continue during the future upgrade of the BESIII experiment.  By the end of the BESIII program, which will include some important machine upgrades,  ten times the current amount  data  will be collected~\cite{Ablikim:2019hff},  and this will usher in  a precision charm flavor era. High statistics data near the production thresholds with quantum-coherent initial states at BESIII will provide key measurements of the phase differences between the decay amplitudes while no reliable QCD-based computation is available. 
 This suite of measurements
is important to the worldwide flavor physics program.
New input from  future BESIII analyses based on larger data samples will  deepen our understanding of the detailed dynamics of charm decays and hopefully facilitate reliable theoretical predictions for the $CP$ asymmetry in charm sector~\cite{Saur:2020rgd}, therefore, allowing us to search for new
physics beyond the SM.

 \begin{table}[htp]
\centering
\caption{\label{tab:prospect}
Prospects of some key measurements with the future data-taking plan in the BESIII white paper~\cite{Ablikim:2019hff}.}
\begin{small}
\begin{tabular}{lccc}
\hline\hline
Observable & Measurement & BESIII~\cite{Ablikim:2019hff}  \\ \hline
$\mathcal B(D^+\to \ell^+\nu_\ell)$&$f_{D^+}|V_{cd}|$&1.1\% \\
$\mathcal B(D^+_s\to \ell^+\nu_\ell)$&$f_{D^+_s}|V_{cs}|$&1.0\%\\
${\rm d}\Gamma(D^{0/+}\to\bar K\ell^+\nu_\ell)/{\rm d} q^2$&$f^K_{+}(0)|V_{cs}|$&0.5\% \\
${\rm d}\Gamma(D^{0/+}\to\pi\ell^+\nu_\ell)/{\rm d} q^2$&$f^\pi_{+}(0)|V_{cd}|$&0.6\% \\
${\rm d}\Gamma(D^+_s\to\eta\ell^+\nu_\ell)/{\rm d} q^2$&$f^\eta_{+}(0)|V_{cs}|$&0.8\%\\
Strong phases in $D^0$&Constraint on $\gamma$ & $<0.4^\circ$   \\
$\Lambda_c^+\to pK^-\pi^+$& $\mathcal B(\Lambda_c^+\to pK^-\pi^+)$ & 2\%   \\
$\Lambda_c^+\to \Lambda \ell^+\nu_\ell$& $\mathcal B(\Lambda_c^+\to \Lambda \ell^+\nu_\ell)$ & 3.3\%  \\
\hline\hline
\end{tabular}
\end{small}
\end{table}

 In addition,  other experiments, such as LHCb and Belle II,  are running and will produce  huge statistics of charm hadrons,  providing  stringent constraints on $CP$-violation
observables~\cite{Saur:2020rgd}. The sensitivity of the observed $CP$ asymmetry in charmed meson decays by LHCb is about $3\times 10^{-4}$~\cite{Aaij:2019kcg} which is consistent with the SM expectation $\mathcal{O}(10^{-4}-10^{-3})$~\cite{Cheng:2012xb}.   BESIII, with 20 fb$^{-1}$ of data at the $\psi(3770)$ peak, can only reach a sensitivity of a few percent level on the $CP$-violation measurements, and the corresponding sensitivity at a super-$\tau$-charm factory~\cite{Bondar:2013cja,Luo:2019xqt} is only $\mathcal{O}(10^{-3})$, which is still one-order-of-magnitude lower than that for the  current LHCb data set~\cite{Xing:2019uzz}.  However,  a super-$\tau$-charm factory has the potential to provide constraints on the decay dynamics of charmed hadrons~\cite{Bondar:2010qs}.
All these experiments plus their future upgrades will continue the studies of charmed hadron physics that will deepen our understanding of  strong interactions in the charm sector, and constrain the SM parameters.  Finally, Table~\ref{tab:prospect} presents the precision prospects for some key charmed hadron measurements that are  based on the BESIII future data-taking plan.   
 
\begin{acknowledgments}
The authors especially thank Prof. S. L. Olsen  for useful comments and suggestion. 
This work is supported in part by the National Key Research and Development Program of China under Contract No. 2020YFA0406400; the National Natural Science Foundation of China under
Contracts Nos. 11822506, 11875054, 11935018; the Chinese Academy of Sciences (CAS) under Contract No. QYZDJ-SSW-SLH003; the CAS Large-Scale Scientific Facility Program; the Fundamental Research Funds for the Central Universities. Conflict of interest statement: none declared.
\end{acknowledgments}

\end{document}